\documentclass[10pt,final, conference]{IEEEtran}
\usepackage{amssymb}
\usepackage{pstricks}
\usepackage{amsmath}
\usepackage[dvips]{graphicx}
\usepackage{enumerate}
\usepackage{ifthen}

\newtheorem{theorem}{Theorem}

\newtheorem{lemma}{Lemma}

\IEEEoverridecommandlockouts

\newcommand{\Ind}[1]{
1_{\left\{#1\right\}}
}

\title{Asymptotic Data Rates of Receive-Diversity Systems with MMSE Estimation and Spatially Correlated Interferers}

\author{\IEEEauthorblockA{Siddhartan Govindasamy, Member}
\thanks{\noindent The author is with the F. W. Olin College of Engineering (siddhartan.govindasamy@olin.edu). This work was supported in part by the National Science Foundation under Grant CCF-1117218. Portions of this work were presented at the 2012 International Symposium on Information Theory (ISIT).}}

\begin{document}

\maketitle
\pagenumbering{arabic}

\begin{abstract}
An asymptotic technique is presented to characterize the data rate
(bits/symbol) achievable on a wireless link in a
spatially distributed network with active interferers at
correlated positions, N receive diversity branches, and linear
Minimum-Mean-Square-Error (MMSE) receivers. This framework is then applied to systems including
 analogs to Matern type I and type II networks which are useful to model systems with Medium-Access Control (MAC), cellular uplinks with orthogonal transmissions and frequency reuse, and  Boolean cluster networks.
It is
found that for our network models, with moderately large N, the
correlation between interferer positions does not
significantly influence the rate resulting in simple
approximations for the data rates achievable in such networks
which are known to be difficult to analyze and for which only a few results are available. These results can help system designers to optimize parameters such as frequency reuse factors in cellular networks and understand the trade off between improved data rates and increased costs associated with increasing diversity orders for wide range of system models.

\end{abstract}


\section{Introduction}
\subsection{Background and Main Contributions}
In systems with multiple diversity branches such as
multiantenna systems, the diversity can be used for interference mitigation
which can increase aggregate data rates in networks.
Since signal and interference strengths are influenced by relative node
positions, the spatial distribution of nodes strongly
influences the performance of such systems. Most works on
wireless networks with spatially distributed nodes assume that
the locations of  active nodes are independent, e.g. by modeling nodes as a Poisson Point Process (PPPs) \cite{JSACPaper} -\cite{TanbourgiIsotropic}.
 The PPP is a good model for networks with spatially uncorrelated transmissions.
In many situations however, active nodes may be spatially correlated, e.g. in networks with
coordinated transmissions such as networks with
Medium-Access-Control (MAC), systems with significant
physical constraints on how close nodes may get to each other
such as vehicular networks, networks which employ location-based spatial reuse such as cellular networks,
and networks with spatially clustered users which naturally arise in many scenarios.

The analysis of networks with spatially correlated transmitters
is  known to be difficult\footnote{We use the term spatially correlated in this work to refer to networks where the \emph{position} in space of randomly distributed transmitters is correlated, as compared to systems where \emph{channel fading} between antennas are correlated}. For instance, Giacomelli, Ganti and Haenggi  remark in \cite{giacomelli2011outage} that ``in view of the difficulties of analyzing
non-Poisson point processes, it cannot be expected that general closed-form expressions exist", and further point out that models that admit repulsion and attraction between nodes are nontrivial. In the uplinks of cellular networks, the spatial dependence of active mobiles due to multiple-access techniques such as Time-Division-Multiple-Access (TDMA) and the dependence of transmit power on position due to power control further complicates analysis, as noted by Novlan, Dhillon and Andrews in \cite{novlan2012analytical}.  Along with   \cite{AndrewsCellular}, \cite{novlan2012analytical} also points out the difficulties in analyzing networks with hexagonal cells which corresponds to an ideal placement of base-stations for covering the plane if coverage regions are circular. To handle the complexity of spatially correlated transmitters, with the notable exception of Poisson cluster networks  (e.g. \cite{GantiClustered} and  \cite{TreschClusteredCell}),  researchers use either asymptotic analysis of some form, or approximate user locations as being independent.   Furthermore, almost all of the existing works deal  with systems that have no diversity (e.g. single antenna users) and thus are unable to exploit diversity for interference mitigation, which as noted earlier is dependent on the spatial distribution of users.

In this paper, we develop an asymptotic technique to characterize the Signal-to-Interference-Ratio (SIR)
of a representative link with multiple receive-diversity branches employing a linear Minimum-Mean-Square-Error (MMSE)
receiver in a network with co-channel interferers at correlated spatial positions. The diversity at the receiver is
used to mitigate interference by optimally combining the signals received on the multiple diversity branches such that the SIR is maximized.  Hence the focus of this work is on the use of diversity to mitigate interference rather than to increase robustness to channel variation which is the focus of a large body of literature on spatial diversity. 
The linear MMSE receiver is the linear receiver that maximizes the SIR if noise is negligible compared to interference. In dense, interference-limited networks the SIR directly influences achievable data rates if Gaussian transmit signals are assumed. Note that if we do not assume Gaussian transmit signals, we can apply a correction factor to the SIR to estimate the rate  \cite{AndrewsCellular}. Additionally, if the residual interference at the output of the MMSE receiver is treated as noise, the Gaussian distribution is the worst case interference distribution if no constraints are placed on the distribution of the transmit signals except for a power constraint. Thus the rate achievable with Gaussian distributed interference can be interpreted as a lower bound on the achievable rate. It is worth noting however, that if more practical constraints are placed on the signal constellation such as M-ary amplitude modulation, the Gaussian distribution is no longer the worst-case distribution \cite{shamai1992worst}. 

We consider the asymptotic regime with the number of diversity branches $N$ and potential interferers $n$ (defined in Section II) going  to infinity with  a positive ratio.   We make this assumption for technical reasons in order to utilize results from infinite random matrix theory that require matrices to grow in both dimensions with a fixed ratio. As we show in Section III, this assumption is not limiting since we later take this ratio to infinity which implies that our results apply to networks with moderately large numbers of diversity branches at the receiver and a large number of active transmitters.  Moreover, the MMSE receiver can null out $N-1$ interferers with $N$ diversity branches. Thus, if $n$ does not grow with $N$, the MMSE receiver will be able to remove all interference resulting in infinite SIR which is not useful in analyzing large, real systems. While the asymptotic analysis provides insight into the scaling behavior of such networks, we use it primarily as a tool to analyze networks with large but fixed numbers of nodes and diversity branches. We do not consider multiple transmit-diversity branches (e.g. multiple transmit antennas) here. In the literature, it has been found that single transmit antennas are often preferable over multiple transmit antennas in dense networks or when low outage probabilities are desired \cite{JSACPaper,jindal2011multi,LouieMultiStream}. Hence only considering single transmit diversity branches is still of interest. We treat the case of  multiple transmit diversity branches in a follow-on work \cite{LimitedRankAsilomar} where we find that in many cases of interest, only one transmit diversity branch should be utilized even when multiple are available.

We find that an appropriately normalized version of the SIR, and with  Gaussian  transmissions, the achievable rate (and its mean), converge in probability to deterministic values if the correlation between active transmitters satisfies an asymptotic independence property which applies to networks where two nodes separated by  large distances are essentially independent. Since it is reasonable to expect nodes that are far away from each other to be independent in many systems, the results here are applicable to a range of systems. The asymptotic expressions for the different spatial correlation models that satisfy our assumptions  have the same form, with the only difference being in the density of active transmitters.   Since networks with spatially independent transmitters are also covered by this model,  the asymptotic rate for systems with ``large" numbers of diversity branches and correlated transmissions is approximated well by the asymptotic rate for a network with independent transmissions but equivalent density of active nodes, resulting in simple approximations, which are exact in an asymptotic sense.

This framework is applied to two hard-core models which are close analogs to Matern type-I and type-II networks commonly used to model Carrier-Sense-Multiple-Access (CSMA), the uplink of cellular networks with large antenna arrays at base stations and hexagonal cells, which is currently an active research area\footnote{While our results  provide insight into achievable rates in \emph{massive} MIMO systems, since our goal is not to analyze such systems specifically, we assume a TDMA is used and do not consider issues such as pilot contamination here. Moreover, our results are applicable even when the number of diversity branches $N$ is only moderately large and not necessarily in the massive regime.}, and a Boolean cluster model where nodes are active if they are within a certain distance of randomly distributed cluster centers.  These examples, which include both repulsive (hard core) and attractive (cluster) models, illustrate the applicability of our  framework to a range of systems. Note that the Boolean cluster model and the Poisson cluster model differ in that in the former, the spatial density of active nodes is either zero or  a positive constant. In the Poisson cluster model however, the density of nodes on the plane could vary significantly as regions where $K > 1$ clusters overlap will have $K$ times the density of nodes compared to regions of the plane where there is a single cluster.

The asymptotic results  are validated by Monte Carlo simulations which show a close agreement between the asymptotic predictions (particularly the mean rate) and simulated values, even when the number of diversity branches per receiver is only moderately large (e.g. $N = 6$). Although six diversity branches (e.g. through six antennas) may be large for certain applications like cellular telephones, it is very realistic for systems such as laptop computers and vehicular communications systems. While it would be ideal to have an exact statistical characterization of the Signal-to-Interference-plus-Noise Ratio (SINR) for finite systems with the forms of spatial correlations that fall under our framework, we note  that such systems have proven  difficult to analyze exactly, even for single antenna systems, as noted in the this and the next subsections,  and evidenced by the approximations that are frequently used to analyze such networks. Thus this work represents and alternative asymptotic approach compared to the technique of taking the density of active interferers to zero utilized in \cite{GantiHighSIR},  and \cite{giacomelli2011outage}.

\subsection{Related Work}
In hard-core networks, active nodes are separated by a minimum distance causing dependence between active nodes which makes these networks particularly difficult to analyze. Bounds on the mean interference in Matern Type-I and II  networks with single-antenna nodes were found by Haenggi in \cite{HaenggiHardCore}. Ganti, Andrews and Haenggi in \cite{GantiHighSIR}, and Giacomelli, Ganti and Haenggi in \cite{giacomelli2011outage}, analyze networks with single-antenna nodes at correlated positions (including hard-core networks) using asymptotic techniques, where the density of active nodes is taken to zero. Recently, El-Sawy et al. \cite{ElSawyCSMA}, \cite{elsawy2013modified}  proposed a new hard-core model for CSMA networks in which they characterize the interference by assuming that active nodes (single antenna) are independently located outside a guard-zone around a test receiver. The prior works that consider \emph{multiantenna} systems in hard-core networks (e.g. \cite{HunterMIMOCSMA}) use a similar approach by approximating  node positions as independent. The uplink of cellular networks with single antenna nodes and Poisson cells were analyzed in \cite{novlan2012analytical} and \cite{NovlanFracReuse}. Cellular systems with regular cells are usually  analyzed by Monte Carlo simulation such as in \cite{XiaoUplinkPowerControl} and other references given in \cite{novlan2012analytical}. In a recent work \cite{CellularNetworks}, we consider the uplink of cellular systems with hexagonal and Poisson distributed cells but with spatially independent wireless nodes. Wireless networks modeled by  Poisson cluster processes have been analyzed in works such as  \cite{GantiClustered} and \cite{TreschClusteredCell}. The high degree of independence in this model, where cluster centers and nodes within each cluster are independently distributed, makes the Poisson cluster model more amenable to analysis than the Boolean cluster model.

\section{System Model}
Consider a circular network of radius $R$ centered at the
origin, with a representative receiver at the origin. The
representative receiver is in a link with a representative
transmitter located at a deterministic point $X_T$ which is at distance
$r_T = |X_T|$ from the origin.  The representative receiver has $N$ diversity branches (e.g. antennas or frequency diversity), with independent flat fading on each diversity branch. In addition to the
representative transmitter, there are $n$  potential
interferers in the network, which if active, are co-channel
interferers to the representative link.
Some of the potential interferers will not
 be actively transmitting as described later in this section.


Let the potential interferer locations be denoted $X_1, X_2,
\cdots, X_{n}$, and $r_i = |X_i|$ be the distance of the $i$-th interferer from the origin. The $n$ potential interferers are i.i.d. with uniform probability in
the circular network such that the area density of potential interferers
$\rho_p$ satisfies
\begin{align}
n = \pi\, \rho_p \, R^2. \label{Eqn:MaternPotentialInterferers}
\end{align}
 The average power (averaged over the fading realizations) received at each diversity branch of the representative receiver from a transmitter at a distance $r_i$, transmitting with power $P_i$ is $P_ir_i^{-\alpha}$, with
$\alpha > 2$. The transmit power of the $i$-th potential interferer, $P_i$
equals one or zero, and the transmit power of the
representative transmitter is fixed at one. Therefore, the
vector of transmit powers $\mathbf{p} = (P_1, P_2, \cdots,
P_n)$ controls which of the potential interferers are active, and the
representative transmitter is assumed to always be active.

The vector $\mathbf{y}\in\mathbb{C}^{N\times 1}$ defined as follows contains the sampled signals at a given sampling time at the $N$
diversity branches of the representative receiver.
\begin{align}
\mathbf{y} = r_T^{-\frac{\alpha}{2}} \mathbf{g}_T  x_T + \sum_{i= 1}^n r_i^{-\frac{\alpha}{2}} \mathbf{g}_i \sqrt{P_i} x_i\,, \label{Eqn:SysEq}
\end{align}
where  $x_i$ ($x_T$), is the  transmitted symbol of  potential interferer $i$ (representative transmitter), and  $\mathbf{g}_i \in \mathbb{C}^{N\times 1}$ ($\mathbf{g}_T \in \mathbb{C}^{N\times 1}$) contains i.i.d., zero-mean,
unit  variance random variables drawn from a continuous distribution representing the fading between the  $i$-th potential interferer (representative transmitter) and the $N$ diversity branches of the representative receiver.
The representative receiver is assumed to perfectly know the channel vector between itself and the representative transmitter $ r_T^{-\frac{\alpha}{2}} \mathbf{g}_T$, and the spatial covariance matrix of the interference,
\begin{align*}
\mathbf{R} = \sum_{i= 1}^n r_i^{-{\alpha}}  {P_i} \mathbf{g}_i\mathbf{g}^\dagger_i\,.
\end{align*}

We shall focus on the interference-limited regime where the noise power $\to$ 0, hence \eqref{Eqn:SysEq} does not include noise.  The potential interferers are assigned
transmit powers of one or zero by a function $g(\cdot)$ such that  $\mathbf{p} =
g\left(X_1, \cdots, X_{n};M_1, M_2,\cdots,M_m;X_T\right)$.
$M_1, \cdots, M_m$ are auxiliary random variables that
are used to select the active transmitters, and are defined differently according to the specific model. For instance, in the HC-II model these variables
are used to select which node is active when there are two or more nodes
within each other's hard core, and in the Boolean cluster model,
these variables represent the cluster centers. The function $g(\cdot)$, thus
controls which of the $n$ potential interferers are transmitting, and since it is parameterized by
$X_T$, the transmit powers of the interferers are a function of the locations of all potential interferers
and the representative transmitter. We denote the set of active interferers as $\mathcal{T} := \{i: P_i =1\}$.

The asymptotic regime we shall consider is the limit as $N$,
$n$ and $R \to \infty$, such that $n/N = c >0$ and $\rho_p$ are
constants, and \eqref{Eqn:MaternPotentialInterferers} holds.
For the rest of this paper,  whenever any one of the quantities $n,N$ or $R \to \infty$, it is assumed that the other two quantities go to infinity as well, such that $n/N= c$  and \eqref{Eqn:MaternPotentialInterferers}  hold.
As mentioned in the introduction, we take $n$ and $N$ to infinity to utilize results from infinite random matrix theory. Since our goal is to analyze large networks with finite densities of users, we need to take $R$ to infinity as $\sqrt{n}$ in order to maintain a fixed density of interferers with increasing $n$. The main results are given in terms of limiting values of a normalized version of
the SIR, $\beta_N = N^{-\alpha/2}r_T^\alpha\mbox{SIR}$, at the
output of the MMSE receiver. This normalization is used because
the (un-normalized) SIR grows with the number of diversity
branches as $N^{\alpha/2}$. Thus, using the standard formula
for the SIR at the output of an MMSE receiver, we have
\begin{align}
\beta_N = N^{-\frac{\alpha}{2}}   \, \mathbf{g}_T^\dagger \, \left(\sum_{i = 1}^n \, P_i\, r_i^{-\alpha}\, \mathbf{g}_i\, \mathbf{g}_i^\dagger\right)^{-1}\mathbf{g}_T\,. \label{Eqn:NormSINRLabel}
\end{align}
Note that the matrix above will be invertible with probability 1 in the limit based on assumptions given subsequently.
Define $p_{in} = N^{\alpha/2}P_i r_i^{-\alpha}$, and  $p_{jn} = N^{\alpha/2}P_j r_j^{-\alpha}$ for convenience.  Note that $p_{in}$ is the average received power (averaged over the fading distribution) from node $i$ scaled by $N^{\alpha/2}$, as observed at an antenna at the origin. We assume that the spatial correlation between the transmitting nodes satisfies the following asymptotic independence property,
\begin{align}
\lim_{n\to\infty}\frac{1}{n^2} \sum_{i= 1}^n \sum_{j = 1}^n &\left[\Pr(p_{in} \leq x, p_{jn} \leq x)\right.\nonumber \\
 &\left.- \Pr(p_{in} \leq x)\Pr(p_{jn} \leq x)\right] = 0\,. \label{Eqn:AsympIndep}
\end{align}
We shall henceforth refer to the asymptotic independence property as AIP.
Note that if  $\Pr(p_{in} \leq x)  =   \Pr(p_{jn} \leq x)\; \forall \, i,j$ and for all $i \neq j$  and $k\neq \ell, \Pr(p_{in} \leq x,p_{jn} \leq x)  =   \Pr(p_{kn} \leq x,p_{\ell n} \leq x)$, which would hold if the pairwise joint CDFs of the $p_{in}$ terms are  equal for each $n$, then \eqref{Eqn:AsympIndep} directly simplifies to the following for $i\neq j$
\begin{align}
\lim_{n\to\infty} \Pr(p_{in} \leq x, p_{jn} \leq x) = \lim_{n\to\infty}\Pr(p_{in} \leq x)\Pr(p_{jn} \leq x)\,. \label{Eqn:AsympIndepSym}
\end{align}
\eqref{Eqn:AsympIndepSym} thus requires that the scaled average received powers at the origin due to two nodes randomly placed in the radius-$R$ network become independent as $R\to\infty$.
We additionally require that
\begin{align}
\lim_{n\to\infty} \Pr(r_i^{-\alpha} N^{\frac{\alpha}{2}} \leq x | P_i = 1) = \lim_{n\to\infty}\Pr(r_i^{-\alpha} N^{\frac{\alpha}{2}} \leq x)\,, \label{Eqn:DistanceFactor}
\end{align}
which requires that the marginal distribution of the scaled path-loss of an active node is independent of whether it is active or not in the limit as $n, N, R\to\infty$.  Additionally, define  $\nu$ as the  average probability of a potential interferer being active in the limit, i.e.,
\begin{align}
\nu = \lim_{n\to\infty}\frac{1}{n}\sum_{i = 1}^n \Pr(P_i = 1)\,.\label{Eqn:NuDef}
\end{align}
We require that ${\nu n}/N = \nu c > 1$, i.e., there are more active interferers than the number of diversity branches at the receiver in the limit. With this requirement, for $n$ and $N$ sufficiently large, there are more active interferers than the number of diversity branches at the receiver and the matrix in  \eqref{Eqn:NormSINRLabel}  consists of a sum of $\nu n$ independent rank 1 matrices. Since $\nu n/N > 1$, and  the entries of $\mathbf{g}_i$ are i.i.d. from a continuous distribution, the matrix in \eqref{Eqn:NormSINRLabel} is full-rank and invertible in the limit with probability 1.

\section{Main Result}\label{Sec:MainResult}
\begin{theorem}
{\it If the system model from the previous section holds, as $n, N, R\to\infty$ such that $n/N = c > 0$, and \eqref{Eqn:MaternPotentialInterferers} hold, $\beta_N \to \beta$ in probability where $\beta$ is the unique, real, positive
solution to
\begin{align}
\frac{2\pi^2\rho \beta^{\frac{2}{\alpha}} }{\alpha}&\csc\left(\frac{2\pi}{\alpha}\right)  =  1 + \frac{2(\pi \rho)^{2-\frac{2}{\alpha}}\beta}{(\alpha -2)(c + \pi \rho \beta)^{1-\frac{2}{\alpha}}}\; \times \nonumber \\
&_2F_1\left(1-\frac{2}{\alpha}, 1-\frac{2}{\alpha}; 2-\frac{2}{\alpha}; \frac{\pi \rho \beta}{ \pi \rho \beta + c}\right)\,, \label{Eqn:LimitingTxCSISINRThinned}
\end{align}
where $\rho = \rho_p \,\nu $, and $_2 F_1(.,.;.;.)$ is the Gauss hypergeometric function (see e.g. \cite{bliss2013adaptive})} 
\end{theorem}
{\it Proof: Given in Appendix \ref{Sec:MainCorrelatedProof}.}

\noindent Note that $\rho$ is  the density of active
interferers in the limit, 
and that \eqref{Eqn:LimitingTxCSISINRThinned} depends on $\rho$, but not on the specific model for the spatial correlation of active transmitters. Since Theorem 1 holds for systems with spatially independent interferers, the normalized SIR in systems with correlated active transmitters (but satisfying the AIP), converges to the same limit as a network with independent transmitters, with the equivalent density of active transmitters. This leads to simple approximations for the rate as the influence on the SIR of the potentially complicated dependence between active nodes becomes negligible with large $N$. For ease of exposition, we shall assume here that the degrees of freedom in the system are spatial degrees of freedom from $N$ antennas at the receiver. The MMSE receiver tends to spatially whiten the interference such that there are no dominant sources of interference in the output of the MMSE receiver. Thus, for large $N$, the residual interference in the output of the MMSE receiver is primarily due to a large number of interferers with approximately the same received power. These interferers occupy a large area, and hence it is the effective density of the active interferers over a large area that primarily influences the interference power, and hence SIR.

Assuming that all transmitters use Gaussian code books and recalling that the noise is negligible, the achievable rate is computed using the Shannon formula, $C_N = \log_2(1+\mbox{SIR})$.  By the continuity of the logarithm, as $N\to\infty$ as in Theorem 1 the following holds in probability.
\begin{align*}
\left|\log_2(1+\mbox{SIR}) - \log_2\left(1+r_T^{-\alpha}N^{\frac{\alpha}{2}}\beta\right) \right| \to 0
\end{align*}
The previous expression implies that the achievable data rate assuming Gaussian transmissions converges to a deterministic asymptote as $N, n, R\to\infty$ in the manner of Theorem 1.  $\beta$ does not depend on the specific spatial correlation model but rather the density of active transmissions. This indicates that features such as imposing a guard-zone in the vicinity of the receiver, do not  influence the achievable data rate beyond reducing the density of active interferers. Note that if we view the $N$ diversity branches as $N$ degrees of freedom available to the MMSE receiver, the MMSE receiver can perfectly null out interference from one dominant interferer at the cost of a one degree of freedom .Thus, when $N$ is large, the performance of systems with or without guard zones (with equal density of active interferers) is approximately equal because the dominant interferers that are removed by the guard-zone in one case, can be nulled with negligible cost in degrees-of-freedom in systems without guard zones.

Furthermore, applying the bounded convergence theorem in the manner of \cite{TxCSIJournal}, we can show that
\begin{align*}
E\left[\log_2(1+\mbox{SIR})\right] - \log_2\left(1+r_T^{-\alpha}N^{\frac{\alpha}{2}}\beta\right)  \to 0.
\end{align*}
Hence, $\log_2\left(1+r_T^{-\alpha}N^{\frac{\alpha}{2}}\beta\right)$ approximates the mean rate for large $N$.

In general,In general, the solution for $\beta$ in \eqref{InitialFixedPoint} has to be found numerically and we haven't been able to prove that numerical techniques that are provably convergent can be used to solve this equation. However, the number of diversity branches at the receiver $N$ is much smaller than the number of potential interferers in the network $n$, $c$ is large, we can find a simple approximation for $\beta$ that exposes the dependency of the SIR on the relevant system parameters. Note that the hypergeometric function in the second term on the RHS of \eqref{Eqn:LimitingTxCSISINRThinned} has positive parameters implying that it is a power series with positive coefficients. Hence, the second term on the RHS of \eqref{Eqn:LimitingTxCSISINRThinned} is non-negative, is an increasing function of its argument, and is bounded from above by
\begin{align}
&\frac{2(\pi \rho)^{2-\frac{2}{\alpha}}\beta}{(\alpha -2)(c + \pi \rho \beta)^{1-\frac{2}{\alpha}}}\; _2F_1\left(1-\frac{2}{\alpha}, 1-\frac{2}{\alpha}; 2-\frac{2}{\alpha}; 1\right) &= \nonumber \\
&\;\;\;\;\;\;\;\;\;\;\;\;\;\;\;\;\;\;\;\;\;\;\;\;\;\frac{2\pi (\pi\rho)^{2-\frac{2}{\alpha}}\beta}{\alpha(c+\pi\rho\beta)^{1-\frac{2}{\alpha}}} \csc\left(\frac{2\pi}{\alpha}\right)\,.
\end{align}
Since $\alpha > 2$,  as $c\to\infty$, the previous expression goes to zero. Thus the second term on the RHS of \eqref{Eqn:LimitingTxCSISINRThinned} goes to zero as well. Thus, if $c\to\infty$  after $n,N,R\to\infty$, we have the following\footnote{We thank Prof. Matthew McKay for suggesting this.}:
\begin{align}
&\beta \to \left[\frac{\alpha}{2\pi^2\rho}\sin\left(\frac{2\pi}{\alpha}\right)\right]^{\frac{\alpha}{2}}\,, \label{Eqn:DoubleLimitSIR} \\
&\lim_{c\to\infty}\lim_{N\to\infty}\left|\log_2(1+\mbox{SIR}) -\right.\nonumber\\
&\left. \log_2\left(\!1\!+\left[\frac{N\alpha}{2\pi^2\rho r_T^2}\!\sin\left(\frac{2\pi}{\alpha}\right)\!\right]^{\frac{\alpha}{2}}\!\right) \right| = 0 \;\text{in probability} \label{Eqn:DoubleLimitSE} \\
&\lim_{c\to\infty}\lim_{N\to\infty}\left|E[\log_2(1+\mbox{SIR})] - \right.\nonumber \\
&\left.\;\;\;\;\;\;\;\log_2\left(1+\left[\frac{N\,\alpha}{2\pi^2\rho r_T^2}\sin\left(\frac{2\pi}{\alpha}\right)\right]^{\frac{\alpha}{2}}\right)\right| = 0. \label{Eqn:DoubleLimitMSE}
\end{align}

Thus for large $N$, but with the number of potential interferers in the network $n$, greatly exceeding the number of diversity branches at the receiver $N$, (i.e. large $c$),
\begin{align}
C\approx E[C]\approx \log_2\left(1+\left[\frac{N\,\alpha}{2\pi^2\rho r_T^2}\sin\left(\frac{2\pi}{\alpha}\right)\right]^{\frac{\alpha}{2}}\right). \label{Eqn:MeanSpecEffApprox}
\end{align}
Since the RHS of \eqref{Eqn:MeanSpecEffApprox} does not depend on the specific value of $c$, this approximation can be used in networks with a fixed but large  $n$ and different values of $N$, as long as $n \gg N$. Variants of this approximation are used in the subsequent sections, and if limits are taken as in \eqref{Eqn:DoubleLimitSE} and \eqref{Eqn:DoubleLimitMSE}, the approximations become precise.

Besides the fact that the achievable rate is not dependent on the specific spatial correlation model but only on the limiting density of active interferers as noted earlier, \eqref{Eqn:DoubleLimitSIR}-\eqref{Eqn:DoubleLimitMSE} provides a simple characterization of the achievable rate as a function of $N$, $\rho$, $\alpha$ and link length $r_T$. Recalling the definition of $\beta_N$, \eqref{Eqn:DoubleLimitSIR} indicates that the SIR grows as $(N/\rho)^{\frac{\alpha}{2}}$ with the MMSE receiver compared to the simpler matched-filter receiver where the SIR grows linearly with $N$. \eqref{Eqn:DoubleLimitMSE} indicates that the mean rate can be kept approximately constant with increasing $\rho$ if $N$ is increased linearly, which suggests that to the extent that the system assumptions hold (in particular the independent fading), a higher density of active transmissions can be accommodated by increasing $N$ e.g. by using a larger number of antennas. Moreover \eqref{Eqn:DoubleLimitSE} indicates that for large $N$ and $c$ and a given link length $r_T$, the achievable rate is approximately constant which is useful for ensuring fairness, in particular for the uplink of power-controlled TDMA cellular networks as described in Section V.

\section{Applications}
\subsection{Hard Core Models}\label{Sec:MaternResult}

We apply this framework to two hard-core models, HC-I and HC-II which are  analogous to the
Matern Type I and II hard-core models \cite{Stoyan}. These models can be used to model active nodes in systems with MAC protocols where nodes that are close to each other are not allowed to transmit simultaneously. For instance, the Matern Type II model has been used to model the CSMA protocol \cite{BaccelliStochasticGeometry1}, which is widely used in wireless networking.

In the HC-I model we assume that all $n$ interferers in the
radius $R$ network are active unless they are within a distance
$h$ of any other  interferer or the representative transmitter.
In the HC-II model, if two (or more) potential interferers are within a distance $h$ of each other,
one of them is activated, resulting in a higher density of active transmitters than in the HC-I model. The auxiliary random variables $M_i$, which are i.i.d. uniform random variables in $(0,1]$, are used to select the active interferer when multiple potential interferers are within a distance $h$ of each other. In the stochastic geometry literature, $M_i$ is known as the \emph{mark} of the $i$-th node. A potential interferer is active only if it has a lower mark than all other potential interferers within a distance $h$ of it.

In comparison, the Matern type-I and II  processes
are constructed from underlying Poisson point processes.
For the Type I process, each point of the
underlying Poisson process within a distance $h$ from
another point of the underlying process is removed. For the
Type II process, a point of the underlying Poisson process is
retained only if there is no other point of the Poisson process
within a distance $h$ of it which has a mark of lesser value.
To incorporate an active transmitter at a particular location, the Matern process has to
be conditioned on having a node at that location which leads to
added mathematical complexity as the probability of a node
being at a particular point is zero. We avoid this
complexity for HC-I and HC-II by assuming that the
representative transmitter is not part of the random process
but is at a deterministic point.

Recall that $(P_1, P_2, \cdots, P_n) = g(X_1, X_2, \cdots, X_n;M_1, M_2, \cdots, M_n;
X_T)$ assigns the transmit powers to the interferers. For the HC-I model we have,
\begin{align}
P_i = \left\{
  \begin{array}{l l}
    0 & \quad \text{if $|X_i -
X_T| < h$ or $\exists \,j$ s.t. $|X_i - X_j| < h$}\\
    1 & \quad \text{otherwise.}
  \end{array} \right.
\end{align}

For the HC-II model,
\begin{align}
P_i = \left\{
  \begin{array}{l l}
    0 & \quad \text{if $|X_i -
X_T| < h$ or  $\exists \,j$  s.t. } \\
\, &\;\;\;\;\;\;\;\;\;\;\;\; \quad |X_i - X_j| < h \text{ and } M_i >
M_j\\
    1 & \quad \text{otherwise.}
  \end{array} \right.
\end{align}
In Appendix \ref{Sec:HardCoreProof}, we show that the HC-I and HC-II models satisfy the AIP in \eqref{Eqn:AsympIndepSym} and \eqref{Eqn:DistanceFactor}. Thus, \eqref{Eqn:MeanSpecEffApprox} can be used to approximate the asymptotic and mean rate with the appropriate limiting density of active transmitters $\rho$. As $R\to\infty$, the limiting densities for the HC-I and HC-II models respectively are found in Appendix \ref{Sec:HardCoreProof} to be
\begin{align}
\rho_{I}  &= \rho_p\mbox{exp}(-\rho_p \pi h^2)\,,\\
\rho_{II} &= \frac{1}{ \pi
h^2}(1-\mbox{exp}(-\rho_p \pi h^2))\,.
\end{align}
The limiting densities above are identical to the density of points in type-I and II Matern hard-core processes, and in conjunction with \eqref{Eqn:MeanSpecEffApprox} can be used to design the radius of guard-zones.

Note that in most works that model hard-core networks in the literature, the authors assume that the users are distributed independently outside the hard-core radius around the representative receiver \cite{ElSawyCSMA, elsawy2013modified,HunterMIMOCSMA}.  This assumption is made {\it a priori} in these works and the theoretical results derived using this assumption are compared to simulations of networks with hard cores around all nodes which introduces correlation in the locations of active users, without a theoretical justification for the approximations.  In contrast we make no such simplifying assumptions in the spatial distribution model we use.

\subsection{Cellular Uplink} \label{Sec:NTDMA}

In certain wireless communications systems such as the uplink of cellular networks with TDMA and frequency reuse, at most one wireless node is allowed to transmit at a given time, in a given frequency band, and in a given spatial region. The active nodes in such a network are spatially correlated and the framework developed here can be used to analyze such systems.

\begin{figure}
\center
\includegraphics[width = 2.15in]{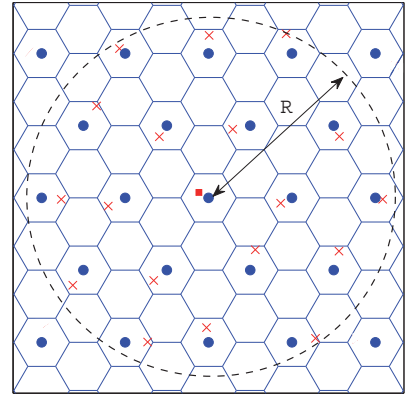}
\caption{Illustration of a cellular network with frequency reuse of 3. The crosses represent mobiles assigned to frequency band 0, and the dots represent base stations assigned to frequency-band 0.}
\label{Fig:Cells}
\end{figure}

Consider the uplink of a cellular network with base stations at hexagonal lattice sites, with one base-station located at the origin acting as the representative receiver which is assigned to frequency band-0. Mobile nodes connect to their nearest base station and so the cells are formed by a Voronoi tessellation of the plane about the base stations.    Let $\rho_c$ denote the area density of the base stations. Overlaid on the grid of base stations is the circular network of radius $R$ centered at the origin, with $n$ mobile nodes, which are the potential interferers, i.i.d. with uniform probability in the circular network as shown in Figure \ref{Fig:Cells} which illustrates the active mobiles in frequency-band 0. Assume that the representative transmitter is in the cell at the origin, at a fixed distance $r_T$ from the origin, and transmits in  frequency-band 0. For simplicity, we shall assume that there is only one time slot in which at most one mobile node per cell is selected to transmit.  Extensions to multiple time slots can be made readily. We shall only consider transmissions in  frequency-band 0 here and all mobiles transmitting in other frequency bands are assumed to transmit with zero power  as they will not influence the signal in frequency-band 0. The positions of the $n$ potential interferers are denoted by $X_1, X_2, \cdots X_n$. As in Section III, we shall take $n, R$ and $N$ to infinity. The $i$-th potential interferer is assigned a mark $M_i$ which is a uniform random variable in $[0,1)$, used to decide which mobile transmits when there are multiple mobiles in a cell assigned to band 0. Let the cells assigned to band 0 be denoted $\mathcal{C}_0, \mathcal{C}_1, \mathcal{C}_2, \cdots$, with $\mathcal{C}_0$ denoting the cell at the origin.

For simplicity of exposition, we focus on systems with unit transmit power at the mobiles, but also provide corresponding expressions for systems that use a simple power control algorithm. The set of active nodes in  band 0  is determined by the function $(P_1, P_2,  \cdots, P_n) = g(X_1, X_2, \cdots, X_n;M_1, M_2, \cdots, M_n;X_T)$ defined as follows.
\begin{align}
P_i = \left\{
  \begin{array}{l l}
    1 & \text{if $\exists \,j$ s.t. $X_i \in \mathcal{C}_j$ and $M_i < M_k\, \forall\, k \neq i$}\\
    \, & \;\;\;\;\;\;\;\;\;\;\;\;\;\;\;\;\;\;\;\;\; \text{ for which $X_k \in \mathcal{C}_j$} \\
    0 &  \text{otherwise}
  \end{array} \right.
  \,.
\end{align}
 Figure \ref{Fig:Cells} illustrates an example of such a network with reuse factor 3, where we have adopted the notation that a reuse factor of $\kappa$ refers to a network where one in $\kappa$ cells are assigned to a given frequency band. The  base stations are represented by the dots and the representative transmitter by the square. The mobile transmitters active in  frequency band 0 are represented by the crosses and the remaining mobiles are not shown.

As outlined in Appendix \ref{Sec:FreqReuseMod}, this model satisfies the AIP in \eqref{Eqn:AsympIndepSym} and
\eqref{Eqn:DistanceFactor}, and the limiting  density of active interferers $\rho$ is found in Appendix \ref{Sec:FreqReuseMod} to be
\begin{align}
\rho = \frac{1}{\kappa}\rho_c\left(1-e^{- \frac{\rho_p}{\rho_c} }\right). \label{Eqn:EffDensNTDMA}
\end{align}
Thus, we can approximate the mean rate of a link with length $r_T$  by \eqref{Eqn:MeanSpecEffApprox} and $\rho$ as above.
If the representative transmitter is located at the cell edge,  $\rho_c = \frac{2}{3\sqrt{3} r_T^2}$. Substituting for $\rho_c$ in \eqref{Eqn:EffDensNTDMA} and the resulting expression for $\rho$ into \eqref{Eqn:MeanSpecEffApprox} yields the following expression for the asymptotic and mean rate of a mobile at a cell edge.
\begin{align}
&C^{\text{asym}}_{\text{pc}} \approx E[C]\nonumber \\
&\;\;\;\;\;\;\approx \log_2\left(1+\left[\frac{N\,\kappa\,\alpha\, 3\sqrt{3}}{4\pi^2\left(1-e^{-\frac{\rho_p}{\rho_c}}\right)}\sin\left(\frac{2\pi}{\alpha}\right)\right]^{\frac{\alpha}{2}}\right). \label{Eqn:MeanSpecEffApproxCellEdge}
\end{align}

Suppose that the $i$-th active mobile transmits with power $r_{bi}^\alpha$ instead of unit power, where $r_{bi}$ is the distance of the $i$-th active mobile node to its base-station. Thus, this power control scheme compensates   for the path loss between the mobiles and their base stations.  Note that this is a special case of the fractional power control scheme from \cite{novlan2012analytical} and others.  Accounting for the power control as in \cite{CellularNetworks}, the asymptotic and mean rate of the representative link is
\begin{align}
&C^{\text{asym}}_{\text{pc}} \approx E[C_{\text{pc}}]\nonumber \\
& \;\;\;\;\approx\log_2\left(1+\left[\frac{ 9\sqrt{3} N\kappa \alpha }{5 \pi^2 \left(1-e^{-\frac{\rho_p}{\rho_c}}\right)}\sin\left(\frac{2\pi}{\alpha}\right)\right]^{\frac{\alpha}{2}}\right). \label{Eqn:MeanSpecEffApproxCellEdgePC}
\end{align}
Thus, for systems with large $N$ and $c$, the achievable rate, regardless of link-length, is close to the RHS of \eqref{Eqn:MeanSpecEffApproxCellEdgePC}.

Assuming the accuracy of the asymptotic approximations, the reuse-normalized asymptotic and   mean rate  can be computed by multiplying the RHS of \eqref{Eqn:MeanSpecEffApproxCellEdgePC}       by $\frac{1}{\kappa}$. Relaxing the integer requirement on $\kappa$, the reuse factor that maximizes the approximate mean rate can be found in terms of the Lambert-W function, $W_0(z)$,  (for details, see Problem 14.6 of \cite{bliss2013adaptive} for a similar calculation in a different context)
\begin{align}
\kappa^* = \left(-\frac{W_0\left(-\frac{\alpha}{2}e^{-\alpha/2}\right)+\alpha}{W_0\left(-\frac{\alpha}{2}e^{-\alpha/2}\right)}\right)^{\frac{2}{\alpha}}\frac{5 \pi^2 \left(1-e^{-\frac{\rho_p}{\rho_c}}\right)}{ 9\sqrt{3} N \alpha }\csc\left(\frac{2\pi}{\alpha}\right). \label{Eqn:OptReuse}
\end{align}
The RHS of \eqref{Eqn:OptReuse} is monotonically decreasing with $\alpha$ and $N$. For $\alpha = 2.5$ and $N = 4$, $\kappa \approx 1$. Since $\alpha \geq 3$ is common in practice, assuming integer reuse, $\kappa = 1$ is optimal for many systems with large $N$.

\subsection{Boolean Cluster Networks} \label{Sec:BoolCluster}


In this section, we illustrate the utility of Theorem 1 to analyze the Boolean cluster model which is described in Section I. More precisely, we can define the Boolean cluster model as follows. Consider a set of cluster centers $M_1, M_2, \cdots M_m$,  i.i.d., with uniform probability, in the circular network such that $m = \pi \rho_b R^2$, where $\rho_b$ is the density of the Boolean clusters. For this model, $(P_1, \cdots, P_n) = g(X_1, X_2, \cdots, X_n;M_1, M_2, \cdots, M_m;
X_T)$ is  as follows.
\begin{align}
P_i = \left\{
  \begin{array}{l l}
    1 & \quad \text{if $X_i \in \bigcup_{i = 1}^m B(M_i, h)$} \\
    0 & \quad \text{otherwise\,,}
  \end{array} \right.
\end{align}
where $B(Y, d)$ is a disk of radius $d$ centered on $Y$. We can show that this model satisfies the AIP because as $R\to\infty$, the probability that two potential interferers $X_i$ and $X_j$ are in the same cluster diminishes to zero. The effective density of active interferers is found to equal the product  of the density of potential interferers $\rho_p$ and the percent coverage of the Boolean model (see e.g. \cite{Stoyan}) with circular grains of radius $h$. This value is
\begin{align}
\rho = \rho_p \left(1-e^{-\rho_b\pi h^2}\right)\,. \label{Eqn:EffDensBool}
\end{align}
Thus, the mean rate is given by \eqref{Eqn:MeanSpecEffApprox} with the above definition for the limiting effective density of active interferers.

\section{Numerical Simulations}\label{Sec:NumSim}
We conducted Monte Carlo simulations to validate  the asymptotic results derived in the previous sections for the different systems considered in this paper.
\subsection{Hard-core models}
We simulated both HC-I and HC-II networks with  $\alpha = 4$. The link-lengths $r_T$ for both sets of simulations were selected such that $\pi \rho_p r_T^2 = 1$, i.e. on average 1 potential interferer is closer to the representative receiver than the representative transmitter. The radius of the guard zones $h$ were varied as a function of $r_T$ and are given in the corresponding figures.

\begin{figure}
\center
\includegraphics[width = 3.15in]{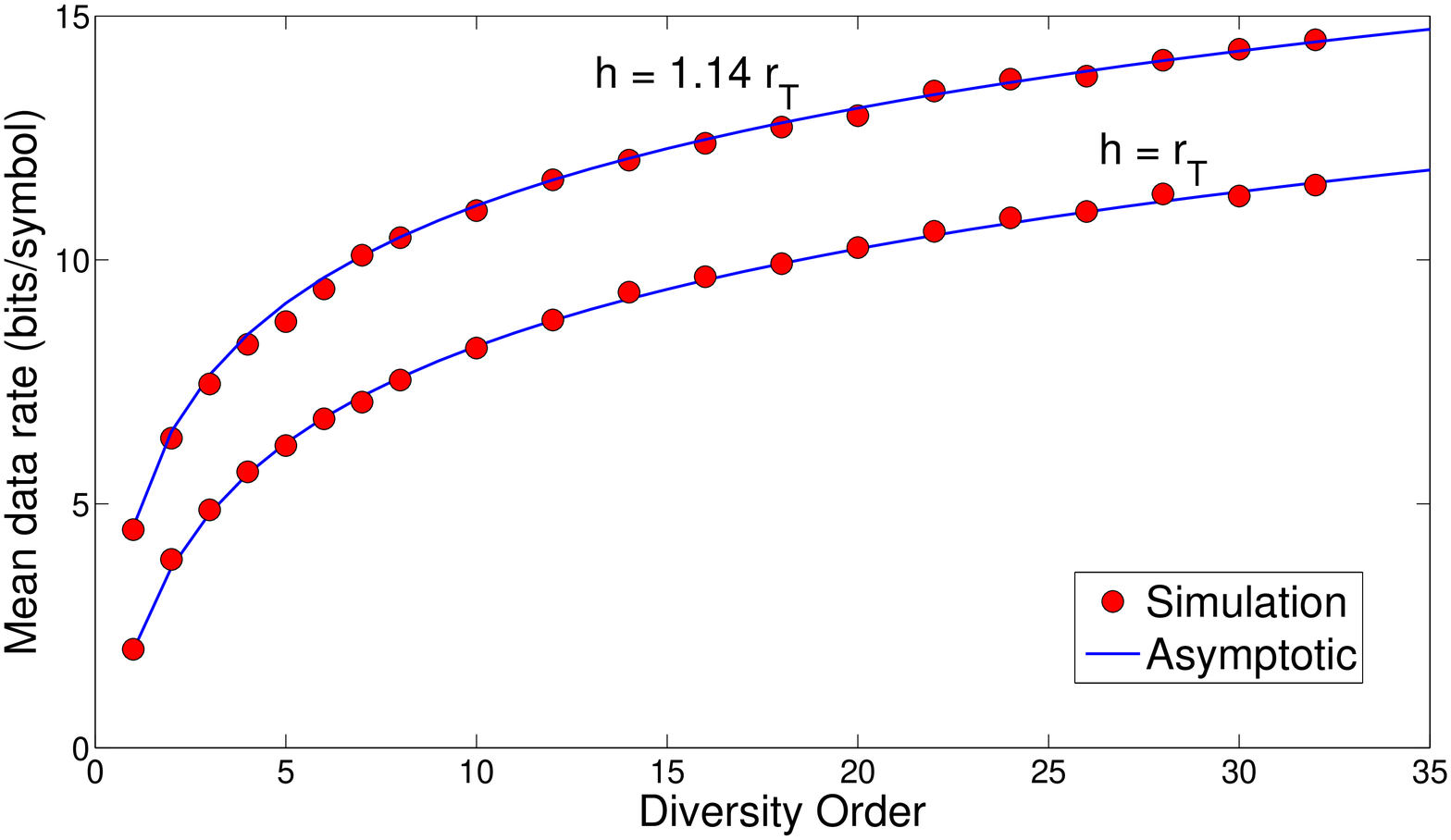}
\caption{Mean rate with the HC-I model of interferers. The link length $r_T$ was such that $\pi\rho_p r_T^2 = 1$, and the radius of the guard zone, $h$ was a multiple of $r_T$ as labeled in the figure.}
\label{Fig:Matern1a4}
\end{figure}

\begin{figure}
\center
\includegraphics[width = 3.15in]{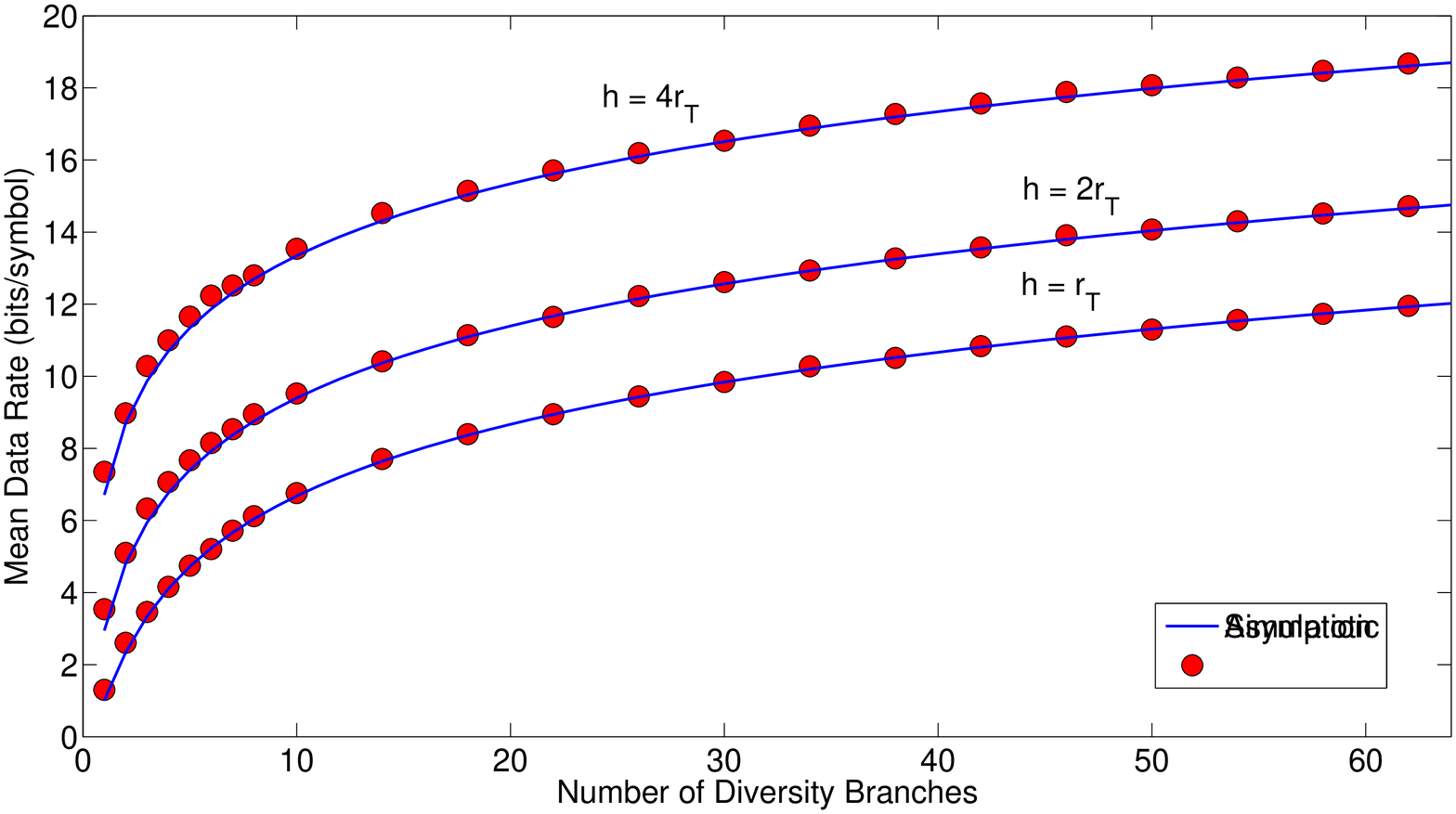}
\caption{Mean rate with HC-II model of interferers.
The link length $r_T$ was such .5that $\pi\rho_p r_T^2 = 1$, and the radius of the guard zone, $h$ was a multiple of $r_T$ as labeled in the figure.}
\label{Fig:Matern2a4}
\end{figure}

\begin{figure}
\center
\includegraphics[width = 3.15in]{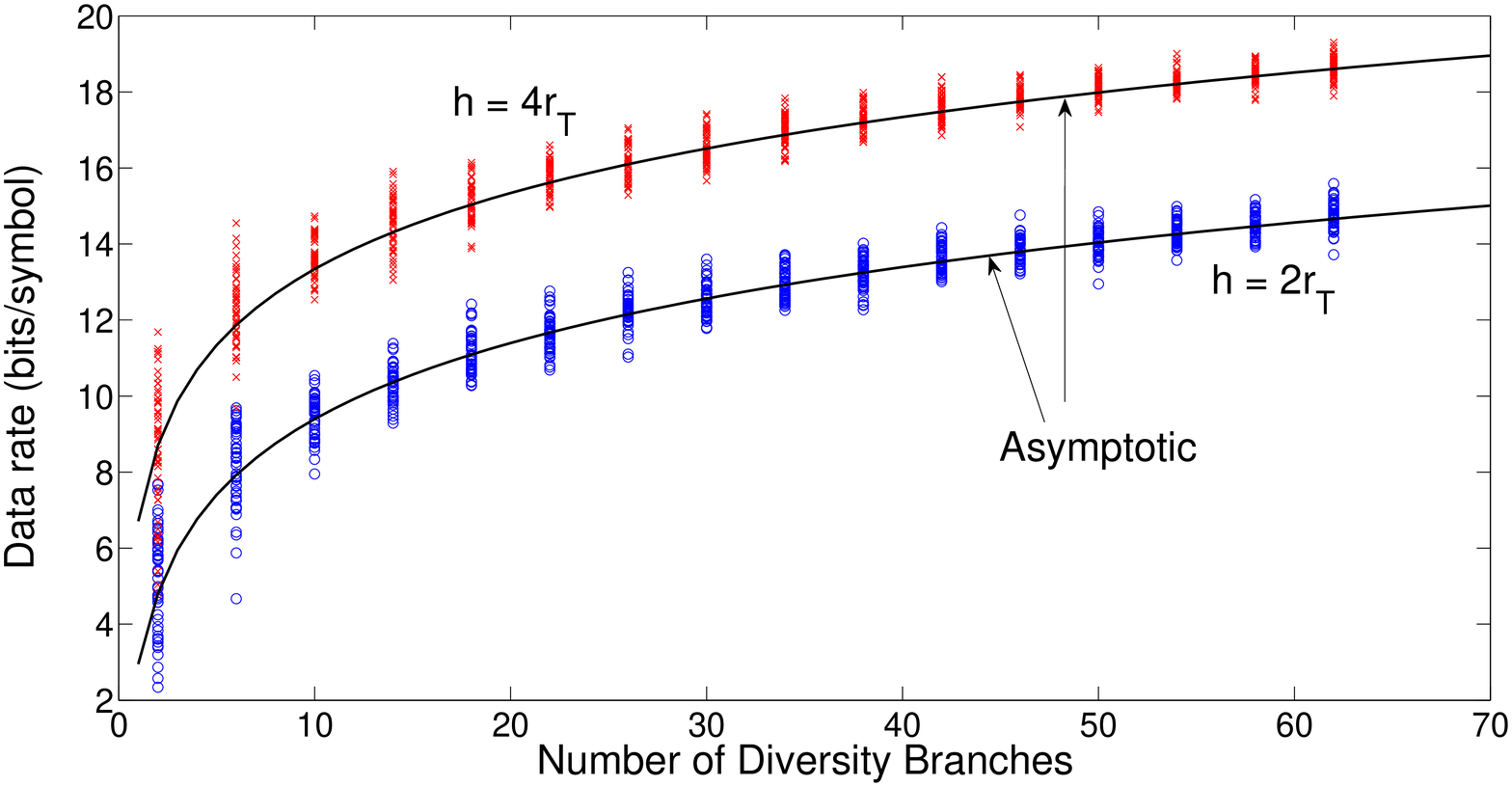}
\caption{Data rate from 100 simulations of HC-II model of interferers vs. number of receive diversity branches. The link length $r_T$ was such that $\pi\rho_p r_T^2 = 1$, and the radius of the guard zone, $h$ was a multiple of $r_T$ as labeled in the figure.}
\label{Fig:Matern2a4Scatter}
\end{figure}

Figure  \ref{Fig:Matern1a4} illustrates the mean rate from 1000 simulations of the HC-I model with guard zones that are multiples of the length of the representative link. As seen in the figure, the asymptotic approximation (from \eqref{Eqn:MeanSpecEffApprox} with $\rho_I$ defined appropriately) is very close to the simulated mean rate.  Figure  \ref{Fig:Matern2a4} illustrates the mean rate from simulations of the HC-II model. Note from the figure that the asymptotic approximation is very close to the simulated mean rate, even when the number of diversity branches is only moderately large. For both the HC-I and HC-II models,
compared to the simulation, for all cases considered the simulated mean rate is within 5\% of the asymptote when $N \geq 2$. Figure  \ref{Fig:Matern2a4Scatter} shows the rate obtained from 100 simulations of the HC-II model for different receive diversity orders to illustrate the distribution of rates. The concentration of points with increasing number of diversity branches indicates that the rate concentrates on the asymptotic prediction which indicates convergence in probability.

\subsection{Cellular Uplink}

\begin{figure}
\center
\includegraphics[width = 3.15in]{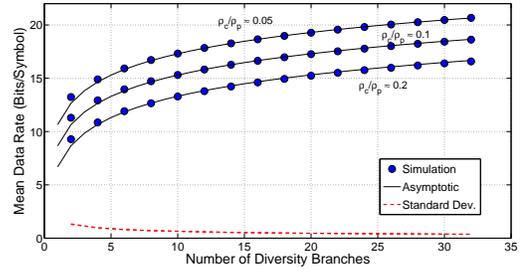}
\caption{Mean rate vs. the number of receive diversity branches for cellular system with reuse factor of 3 and different relative densities of wireless nodes to base stations. Representative link length $r_T$ is such that $\pi \rho_p r_T^2 = 1$. The three dashed lines (difficult to distinguish on this scale) represent the decaying standard deviation of the rate. }
\label{Fig:NTDMARank1}
\end{figure}

To validate the asymptotic results presented in Section \ref{Sec:NTDMA}, we simulated cellular networks with hexagonal cells and frequency reuse factor 3 as illustrated in Figure \ref{Fig:Cells}. The path-loss exponent  $\alpha = 4$, the density of potential interferers (mobiles) $\rho_p = 0.01$ and different relative densities of base stations to mobile nodes (indicated in the Figure) were used. At most one mobile was allowed to transmit  in any cell assigned to band 0. Figure \ref{Fig:NTDMARank1} illustrates the mean rate of the representative link of length $r_T$, where $r_T$ was such that $\pi \rho_p r_T^2 = 1$, i.e. there is on average one potential interferer closer to the base station at the origin than the representative transmitter. The representative transmitter was assigned to band 0 and is always active.

The markers in Figure \ref{Fig:NTDMARank1} illustrate the simulated values and the solid lines represent \eqref{Eqn:MeanSpecEffApprox}, with the effective density of interferers $\rho$ given by \eqref{Eqn:EffDensNTDMA}. The asymptotic prediction for the mean rate are within 1\% of the simulated values for $N \geq 6$, validating the asymptotic predictions. The dashed lines represent the simulated standard deviations which decay with number of diversity branches. For the systems we simulated, the standard deviation is less than 5\% of the asymptotic mean rate when $N \geq 10$. The decaying standard deviation indicates mean-square convergence which implies convergence in probability.

\begin{figure}
\center
\includegraphics[width = 3.45in]{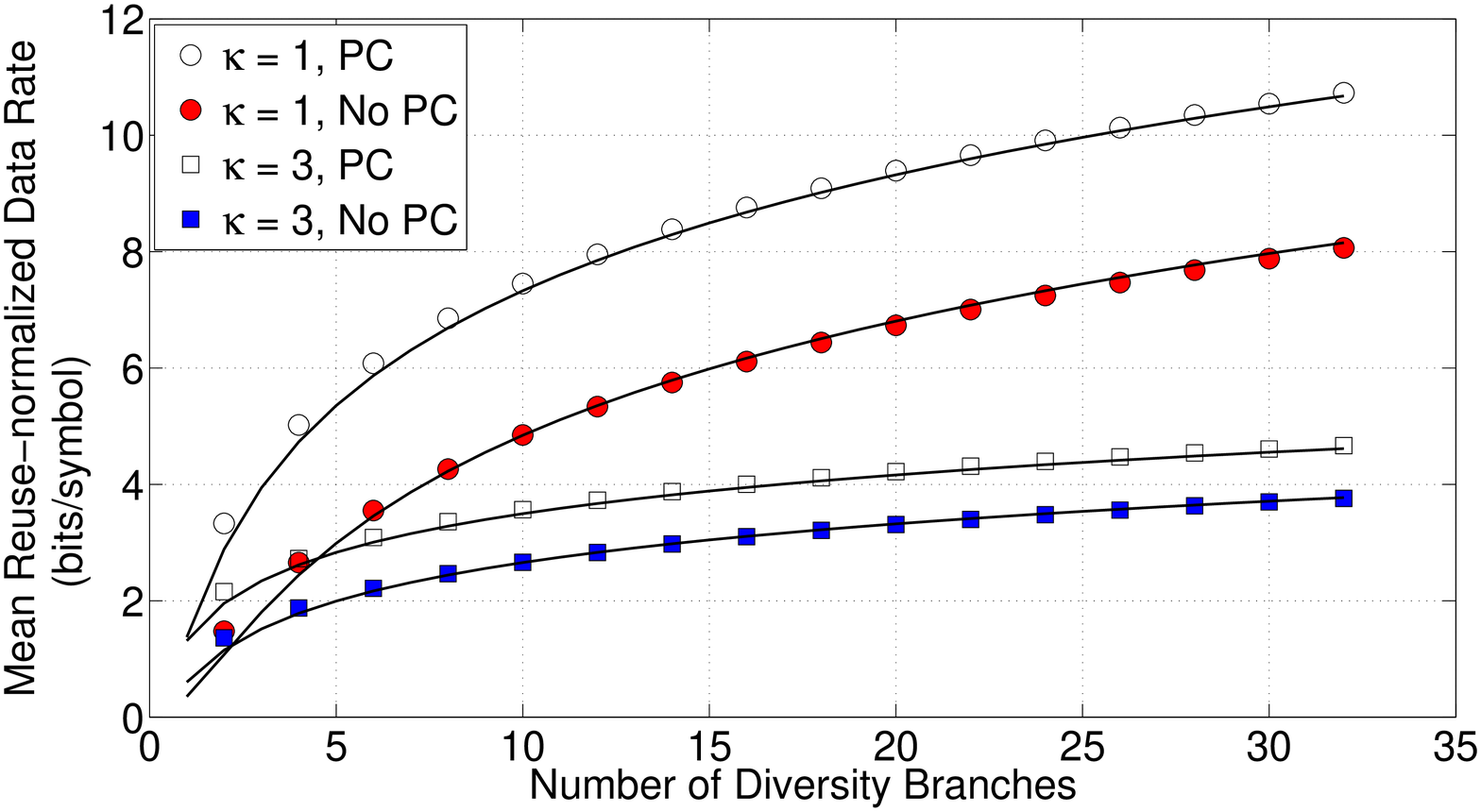}
\caption{Mean rate normalized by reuse factor vs. the number of receive diversity branches for the uplink of a TDMA cellular system with reuse factors of $\kappa=3, 1$, $\rho_c/\rho_p = 0.1$, with and without power control. The representative transmitter is located at the edge of the cell.}
\label{Fig:NTDMACellEdge}
\end{figure}

Figure \ref{Fig:NTDMACellEdge} illustrates simulations of a  network with $\rho_c/\rho_p = 0.1$, with the representative transmitter located at the edge of the cell centered at the origin. The figure shows the reuse-normalized mean rate vs. the number of diversity branches at the representative base station for systems with and without power control, and reuse factors of $\kappa = 1$ and $3$. Note that power control can increase the mean rate of the cell-edge user by approximately 40\% and that a frequency reuse factor of 1 results in greater reuse-normalized mean rate for power controlled systems than a reuse factor of 3.

\subsection{Boolean Cluster Model}
\begin{figure}
\center
\includegraphics[width = 3.15in]{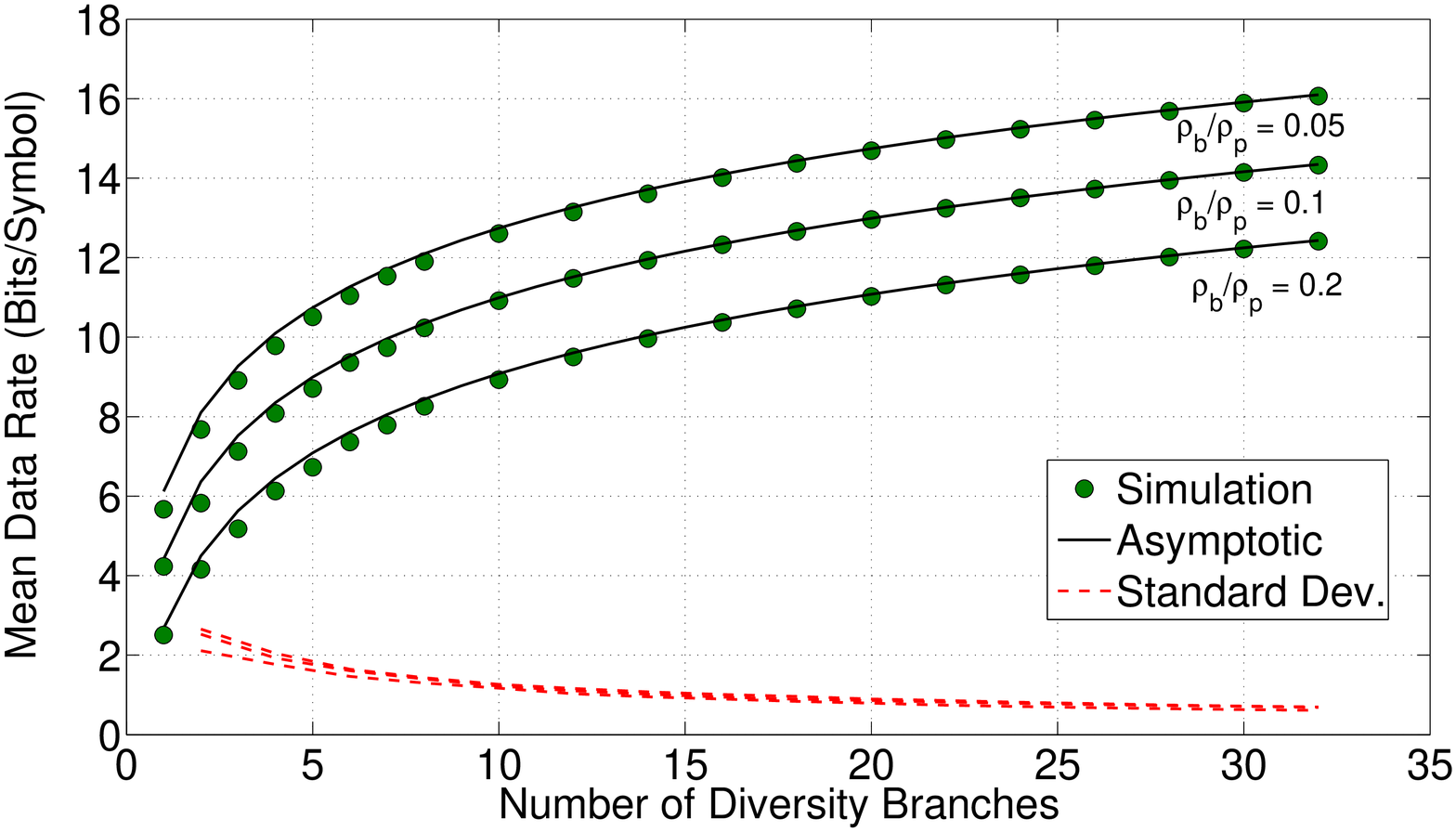}
\caption{Mean rate vs. number of diversity branches for Boolean cluster model with varying cluster density. The three dashed lines represent the decaying standard deviation of the rate. }
\label{Fig:BoolCluster}
\end{figure}

We simulated Boolean cluster networks with path-loss exponent  $\alpha = 4$. The density of potential interferers $\rho_p = 0.01$ and different relative densities of clusters to wireless nodes was used. Figure \ref{Fig:BoolCluster} illustrates the mean rate of the representative link of length $r_T$. $r_T$ was such that $\pi \rho_p r_T^2 = 1$, i.e. there is on average one potential interferer closer to the representative base station than the representative transmitter.
The simulated mean rate are represented by the markers, and the asymptotic prediction of \eqref{Eqn:MeanSpecEffApprox} with the effective density of active interferers given by \eqref{Eqn:EffDensBool}, is represented by the solid line. The simulated mean rate agrees with the asymptotic expression to within 2.5\% when $N\geq 8$, and within 5\% if $N\geq 6$ in all the cases we considered. The dashed lines in the figure indicate the standard deviation of the  rate from simulations which decay with increasing  numbers of diversity branches. For the cases we consider, the standard deviation is 10\% or less of the asymptotic mean rate when the number of diversity branches is 12 or more. The decay in standard deviation indicates convergence in mean square, and hence in probability.

\section{Summary and Conclusions}

An asymptotic technique to characterize the rate achievable
in wireless networks where active interferers are spatially
correlated and receivers have $N$ diversity branches with
linear MMSE processing is presented.  We show that for large $N$ and assuming Gaussian transmissions, the rate and its mean approach an asymptote. While systems with spatially correlated transmitters are known to be difficult to analyze, in networks that satisfy our assumptions, when $N$ grows large the influence of the spatial dependence of the active nodes on the rate diminishes. Thus, the rate and in particular its mean can be approximated using simple expressions involving the limiting density of active transmissions,  which are precise in an asymptotic sense.

We consider networks where the path-losses from active transmitters satisfy a certain pairwise asymptotic independence criteria. This framework is applied to two hard-core models,  which are close analogs to Matern type I and II models which are commonly used to model networks with CSMA MAC protocols. We also apply this result to analyze the uplink in cellular networks with orthogonal transmissions and frequency reuse, and a Boolean cluster model which can be used to model networks with hot spots.

These findings help improve our understanding of the tradeoff between increasing numbers of diversity branches, which is directly related to cost in most systems, on  data rates. These results can also be utilized to optimize parameters such as reuse factors in cellular systems.  The asymptotic results are validated by Monte Carlo simulations which match theoretical predictions even when the number of diversity branches is only moderately large. These findings provide insight into wireless networks with correlated transmissions, which are well known to be challenging to analyze, and for which only a limited number of results exist in the literature.

\section{Acknowledgements}
We thank J. Zhu and Profs. M. McKay and R. Louie for helpful comments and discussions. We also thank the anonymous reviewers for their helpful comments.

\appendix
\subsection{Proof of Main Result} \label{Sec:MainCorrelatedProof}
To prove the main result, we first define the empirical distribution function (e.d.f.) of a set of
random variables as the proportion of those random
variables that are less than or equal to the argument of the
e.d.f. For instance, denoting the e.d.f. of  $p_{in}$ by $H_n(x)$, we have
\begin{align}
H_n(x) = \frac1n\sum_{i = 1}^n 1_{\{p_{in}  \leq x \}}= \frac1n\sum_{i = 1}^n a_{in}(x)\,,\label{eqn19}
\end{align}
where $1_{\{\cdot\}}$ is the indicator function, and  $a_{in}(x) = 1_{\{p_{in} \leq x\}}$.
The main property used in proving Theorem 1 is the following
lemma which is a variant of Lemma 1 from Govindasamy et al. \cite{CellularNetworks} and Lemma 4.3 of Tse and Hanly
\cite{TseMIMOCDMA}. The  difference between this lemma and Lemma 1 from \cite{CellularNetworks} is that in \cite{CellularNetworks}, the e.d.f. of the $\psi_i$ terms  is assumed to converge with probability 1.  This property holds by the strong law of large numbers if the $\psi_i$ terms are independent (corresponding to spatially independent transmitters). In this work, the convergence is only required to be in probability which enables us to apply the lemma to a wider class of models with spatially correlated active transmitters.
\begin{lemma} \label{Lemma:SIRConvergenceLemma}
Consider the quantity
\begin{align}
\gamma_N = \frac1N \mathbf{s}^\dagger\left(\frac1N\mathbf{S \Psi S}^\dagger\right)^{-1}\mathbf{s} \label{eqn20}
\end{align}
where $\mathbf{s} \in \mathbb{C}^{N\times 1}$ and $\mathbf{S}
\in \mathbb{C}^{N\times n}$ comprise i.i.d., zero-mean,
unit-variance entries, $n/N = c$, and  $\mathbf{\Psi} =
\mbox{diag}(\psi_1, \psi_2, \cdots \psi_{n})$. Suppose that as
$n, N \to\infty$, the e.d.f. of $(\psi_1, \psi_2, \cdots \psi_{n})$ converges in probability to  $H(x)$.
Additionally, assume that there exists an $n_0$ such that
$\forall n > n_0$, $\lambda_{min}\left(\frac1N\mathbf{S \Psi S}^\dagger\right) \geq \lambda_{\ell b} > 0$, with probability 1, where $\lambda_{min}(\mathbf{A})$  denotes the minimum eigenvalue of any square matrix $\mathbf{A}$, and $\lambda_{\ell b}$ is an arbitrary, strictly positive number. Then,
$\gamma_N \to \gamma$ in probability where $\gamma$ equals
the unique non-negative real solution for $\gamma$ in
\begin{equation}
1 =
\gamma\,c\int_{0}^{\infty} \frac{\tau dH(\tau)}{1+\tau \gamma}\,.\label{InitialFixedPoint}
\end{equation}
\end{lemma}
\noindent \emph{Proof: Given in Appendix \ref{Lemma:SIRConvergenceLemmaProof}. }

Let  $\mathbf{s} = \mathbf{g}_T$,
$\mathbf{\Psi} = diag\left(N^{\alpha/2} P_1 r_1^{-\alpha},N^{\alpha/2}
P_2r_2^{-\alpha}, \cdots\!, N^{\alpha/2} P_n r_n^{-\alpha}\right)$,
and $\mathbf{S} = \left[\mathbf{g}_1\,\mathbf{g}_2 \, \cdots
\,\mathbf{g}_n\right]$. Comparing with \eqref{Eqn:NormSINRLabel}, we note that $\beta_N = \gamma_N$ from \eqref{eqn20}. To use Lemma \ref{Lemma:SIRConvergenceLemma} to characterize $\beta_N$, we first condition on  the event (denoted by $\mathcal{D}$) that the fraction of potential interferers that are active is greater than or equal to $\frac{1}{\delta}$ of its limiting value. Thus, $\mathcal{D}$ is the event that $\frac{|\mathcal{T}|}{n} \geq \frac{\nu}{\delta}$, which from the definition of $c$ from Section II implies that $|\mathcal{T}| > N$. 
This condition implies that for $N$ sufficiently large,  $\frac1N\mathbf{S \Psi S}^\dagger$ is full rank, is invertible and its minimum eigenvalue is strictly positive, satisfying the minimum eigenvalue condition required by Lemma 1, for the same reasons that the matrix in \eqref{Eqn:NormSINRLabel} is invertible with probability 1 in Section II.
Since $\frac{|\mathcal{T}|}{n}\to\nu$ in probability by the AIP, as $n\to\infty$, $\Pr\left(\nu -\frac{ |\mathcal{T}|}{n} > \nu-\frac{\nu}{\delta}\right) \to 0$, implying 
\begin{align}
\Pr\left(\mathcal{D}\right) = \Pr\left(\frac{|\mathcal{T}|}{n} \geq \frac{\nu}{\delta}\right)  \to 1\,. \label{Eqn:ActiveFractionMatConv}
\end{align}
To use Lemma \ref{Lemma:SIRConvergenceLemma}, we first show
 that conditioned on $\mathcal{D}$, the e.d.f. of the $p_{in}=N^{\alpha/2}P_ir_i^{-\alpha}$ terms, defined in \eqref{eqn19}, converges  in probability to a limiting function $H(x)$ in Lemma \ref{Lemma:LimitingEDF}. We then show that the minimum
eigenvalue property required by Lemma \ref{Lemma:SIRConvergenceLemma} is satisfied in Lemma \ref{Lemma:SmallestEval}.  The proof is completed by evaluating \eqref{InitialFixedPoint} with $H(x)$ from Lemma \ref{Lemma:LimitingEDF}.

\begin{lemma} \label{Lemma:LimitingEDF}
If \eqref{Eqn:AsympIndep} and \eqref{Eqn:DistanceFactor} hold, and conditioned on $\mathcal{D}$, as $n, N, R \rightarrow \infty$,
\begin{align}
H_n(x) \to H(x)=1-\frac{\pi\rho}{c}x^{-\frac{2}{\alpha}}\Ind{\left(\frac{\pi\rho_p}{c}\right)^{\alpha/2}
< x }\,, \label{eqn24}
\end{align}
in probability.

\noindent \emph{Proof: Given in Appendix
\ref{Sec:ProofOfLimitingEDF}}
\end{lemma}

\begin{lemma} \label{Lemma:SmallestEval}
Let $\mathbf{\Psi}$ and $\mathbf{S}$ be as defined above and recall that $1< \delta < c\nu$. Conditioned on $\mathcal{D}$ there exists an $n_0$
such that for all $n > n_0$, and with probability 1,
\begin{align}
\lambda_{min}\left(\frac1N\mathbf{S \Psi S}^\dagger\right) &= \lambda_{min}\left(\frac1N \sum_{i = 1}^nN^{\alpha/2}r_i^{-\alpha} P_i \mathbf{g}_i\mathbf{g}_i^\dagger\right)\nonumber \\ &>  \left(\frac{\pi \rho_p}{c}\right)^{\frac{\alpha}{2}}\frac{c\nu}{2\delta} \, \left(1-\sqrt{\frac{\delta}{c\nu}}\right)^2\,. \label{Eqn:MinEvalEqnDef}
 \end{align}

\noindent \emph{Proof: Given in Appendix \ref{Sec:ProofOfSmallestEval}}
\end{lemma}
Thus,  from Lemmas \ref{Lemma:LimitingEDF}, \ref{Lemma:SmallestEval} and   \ref{Lemma:SIRConvergenceLemma},  conditioned on $\mathcal{D}$, $\beta_N$ converges to a limit  we define as $\beta= \gamma$ in probability. Hence, for any $\epsilon > 0, \Pr\left(|\beta_N-\beta| > \epsilon\left|\mathcal{D}\right.\right)\to 0\,$. Since $\Pr(\mathcal{D}) \to 1$, for any $\epsilon > 0$,
\begin{align}
\Pr\left(\left|\beta_N-\beta\right| > \epsilon\right)\to 0\,,
\end{align}
i.e. $\beta_N \to\beta$ in probability. We can then substitute the derivative of $H(x)$ from \eqref{eqn24},
into \eqref{InitialFixedPoint} (with $\beta$ now replacing $\gamma$), and integrate to get
\begin{align}
\frac{2\pi^2\rho\beta^{\frac{2}{\alpha}}}{\alpha}& {\rm
csc}\left(\frac{2\pi}{\alpha}\right)-\frac{2(\pi\rho)^{2-{\frac{2}{\alpha}}}\beta}{(\alpha-2)c^{1-\frac{2}{\alpha}}}\nonumber \\
&{_2}F_1\left(1,1-\frac{2}{\alpha},2-\frac{2}{\alpha};-\frac{\pi\rho\beta}{c}\right)=1\,.\label{eqn27}
\end{align}
Applying the Pfaff transform (see e.g. \cite{bliss2013adaptive}) to \eqref{eqn27} yields \eqref{Eqn:LimitingTxCSISINRThinned}.

\subsection{Proof Of Lemma \ref{Lemma:SIRConvergenceLemma} } \label{Lemma:SIRConvergenceLemmaProof}
Let $Q_N(x)$ denote the e.d.f. of the diagonal entries of $\mathbf{\Psi}$, and $Q(x)$ be its limit as $N, n, R\to\infty$ in the manner of Lemma 1 (note that all convergence results here refer to this regime). Additionally, let $G_N(\cdot)$ denote the e.d.f. of the eigenvalues of $\frac1N\mathbf{S \Psi S}^\dagger$ and $G(\cdot)= \lim_{N\to\infty}G_N(\cdot)$ which exists as shown subsequently. The proof is based on relaxing the requirement that $Q_N(x) \to Q(x)$ with probability 1 from Lemma 1 of \cite{CellularNetworks}, to convergence in probability.  If  $Q_N(x)$ converges with \emph{probability 1} to $Q(x)$, from Theorem 4.3 of \cite{bai2009spectral}, $G_N(x) \to G(x)$ with probability 1 . The proofs of  Lemma 1 from \cite{CellularNetworks} and the corresponding result from \cite{TseMIMOCDMA}  use  this fact to show that the following  converges to zero.
\begin{align}
\sum_{k = 1}^L \Pr\left(\left|G_N(I_k) - G(I_k)\right|> \xi/L\right)\,, \label{Eqn:EDFPartitionEqn}
\end{align}
where $I_k$ is the $k$-th interval of a partition of the interval $(\lambda_{\ell b}, \infty)$ into $L$ intervals, and $\xi$ is an arbitrary positive number. The expressions $G_N(I_k)$ and $G(I_k)$ should be interpreted respectively as the probability that a random variable with CDF $G_N(\cdot)$ or $G(\cdot)$ falls in the interval $I_k$. From the definition of convergence in probability, \eqref{Eqn:EDFPartitionEqn} goes to zero even if $G_N(x)\to G(x)$  only in probability.  Since this is the only place in which the convergence with probability 1 of $G_N(x)$ to $G(x)$ is used in the proof of  Lemma 1 from \cite{CellularNetworks}, Lemma 1 in this paper can be proved by showing that  $G_N(x) \to G(x)$ in probability if $Q_N(x) \to Q(x)$ in probability. To prove this, we invoke an argument used to prove Theorem 4.1 of Bai and Silverstein \cite{bai2009spectral}. From Theorem 4.3 of \cite{bai2009spectral}, as $N\to\infty$ if $Q_N(x) \to Q(x)$ with probability 1, $G_N(x) \to G(x)$ with probability 1, and from Theorem 4.1 of Bai and Silverstein \cite{bai2009spectral} if  $Q_N(x)\to Q(x)$ in probability, then $G_N(x)$ converges to \emph{some} limit $\bar{G}(x)$ in probability (no form for $\bar{G}(x)$ is given in that theorem).  The remaining task is to show that $\bar{G}(x) = G(x)$. For a given $x$, consider the pair of random variables $(G_N(x), Q_N(x))$. By Theorem 4.3 of \cite{bai2009spectral}, if   $Q_N(x)\to Q(x)$ with probability 1, the pair $(G_N(x), Q_N(x))$ converges with probability 1 to the pair $(G(x), Q(x))$. By the Skorohod representation theorem (see e.g. \cite{Karr}),  there exists a probability space in which the pair $(G_N(x), Q_N(x))$ converges in probability but not with probability 1 to the pair $(G(x), Q(x))$. Since we know that if $Q_N(x)\to Q(x)$ in probability, $G_N(x)\to\bar{G}(x)$ in probability, it must be the case that  $\bar{G}(x)=G(x)$. We have thus relaxed the requirement on the convergence of $Q_N(x) \to Q(x)$, proving the main part of the lemma. Finally, the uniqueness of the solution for $\gamma$ follows directly from Proposition 3.2 of  \cite{TseMIMOCDMA} which shows that an equation of the form in \eqref{InitialFixedPoint} has only one solution for $\gamma$ in $[0, \infty)$.

\subsection{Proof of Lemma
\ref{Lemma:LimitingEDF}}\label{Sec:ProofOfLimitingEDF}
We first show that this property holds without conditioning on $\mathcal{D}$.
From   the definition of  $a_{in}(x)$, $E[a_{in}(x)]={\rm{Pr}}(p_{in}\leq x)$ and
$E[a_{in}(x) a_{jn}(x)] = {\rm{Pr}}(p_{in}\leq x,p_{jn}\leq
x)$. Next, define
\begin{align}
&\bar{H}_n(x) = \frac{1}{n} \sum^n_{i=1}
E[a_{in}(x)]=\frac{1}{n}\sum^n_{i=1} {\rm{Pr}}
\big(r_i^{-\alpha} P_i N^{\alpha/2} \leq x\big)\notag\\
&=\frac{1}{n}\sum_{i =
1}^n[\Pr\left(N^{\frac{\alpha}{2}} r_i^{-\alpha}
\leq x |P_i=1 \right)\! \Pr(P_i = 1)\nonumber \\
&\;\;\;\;\;\;\;\;\;\;+\Pr\left(P_iN^{\frac{\alpha}{2}} r_i^{-\alpha}
\leq x |P_i=0 \right)\!\Pr(P_i = 0)]\label{Eqn:BarHDef}
\end{align}
where  \eqref{Eqn:BarHDef} follows from the fact that $\Pr\left(P_iN^{\frac{\alpha}{2}} r_i^{-\alpha}
\leq x |P_i=0 \right) = 1$ for $x \geq 0$.
Note from \cite{JSACPaper} that the CDF of
$N^{\alpha/2}r_i^{-\alpha}$ is
\begin{align}
\Pr(N^{\alpha/2}r_i^{-\alpha} \leq x) =
1-\frac{\pi\rho_p}{c}x^{-\frac{2}{\alpha}}\Ind{\left(\frac{\pi\rho_p}{c}\right)^{\alpha/2}
< x }\,,\label{eqn36}
\end{align}
which does not depend on $N$ because $N$,  $n$, and $R$ are related such that the
CDF of $N^{\alpha/2}r_1^{-\alpha}$ is independent of $N$. Taking the limit of \eqref{Eqn:BarHDef}  and substituting \eqref{Eqn:DistanceFactor} and \eqref{eqn36} yields
\begin{align}
&\lim_{n\to\infty}\bar{H}(x) = \lim_{n\to\infty} \frac{1}{n}\sum_{i = 1}^n \Pr(P_i =
1) \;\times \nonumber \\
&\left(1-\frac{\pi\rho_p}{c}x^{-\frac{2}{\alpha}}1_{\{\left(\frac{\pi\rho_p}{c}\right)^{\alpha/2}< x\}}\right)+\lim_{n\to\infty}
\frac{1}{n}\sum_{i = 1}^n \Pr(P_i = 0).\notag\\
&=H(x)\,. \label{Eqn:BarHnLim}
\end{align}
where the last equality follows from substituting
$\rho = \left(\lim_{n\to\infty} \frac{1}{n}\sum^n_{i=1}\Pr(P_i = 1
)\right) \rho_p$ and the fact that $\Pr(P_i = 1) + \Pr(P_i = 0) = 1$. What remains now is to show that $H_n(x)\to H(x)$ in probability.

From \eqref{Eqn:BarHnLim}, for each $\epsilon>0$, $\exists \, n_a$ such that for all $n > n_a$, $ |\bar{H}_n(x)- H(x)| <  \frac12\epsilon$. Thus $\forall\, n > n_a$,
\begin{align}
&\Pr\left(\left|H_n(x)-H(x)\right| > \epsilon \right) \leq \Pr\left(\left|H_n(x) - \bar{H}_n(x)\right| \geq \frac{\epsilon}{2} \right) \nonumber \\
&=\Pr\left(\left|\sum_{i=1}^n a_{in}(x) - \sum_{i=1}^n
E[a_{in}(x)]\right| \geq \frac{n\epsilon}{2} \right)\nonumber\\
& \leq
\frac{4}{\epsilon^2 n^2}\mbox{var}\left\{\sum_{i=1}^n
a_{in}(x)\right\}\nonumber\\
&=\frac{4}{\epsilon^2 n^2}
\sum_{i=1}^n \sum_{j=1}^n {\rm cov} (a_{in}(x),a_{jn}(x))\nonumber \\
&=\frac{4}{\epsilon^2 n^2}\sum_{i=1}^n \sum_{j=1}^n
(E[a_{in}(x) a_{jn}(x)]-E[a_{in}(x)]E[a_{jn}(x)])\nonumber\\
&=\frac{4}{\epsilon^2 n^2} \sum_{i=1}^n \sum_{j=1}^n
\left({\rm{Pr}}(p_{in}\leq x,p_{jn}\leq x)\right.\nonumber \\
&\;\;\;\;\;\;\;\;\;\;\;\;\;\;\;\;\;\;\;\;\;\;\;\;\;\;\;\;\;\;\;\;\;\;\;\;\left.-{\rm{Pr}}(p_{in}\leq x)
{\rm Pr}(p_{jn}\leq x)\right),\label{eqn32}
\end{align}
where the inequality follows from the Chebyschev inequality, and the subsequent equation from the expansion of the variance of a sum of random variables. As $n,N, R\to\infty$, from \eqref{Eqn:AsympIndep} we have the RHS of \eqref{eqn32} $\to 0$ which implies that $H_n(x)\to H(x)$ in probability. Since $\Pr(\mathcal{D})\to 1$, this holds when conditioned on $\mathcal{D}$ as well.

\subsection{Proof of Lemma \ref{Lemma:SmallestEval}}\label{Sec:ProofOfSmallestEval}
Recall that $\mathcal{T}$ is the set of active transmitters, and the definitions of $\mathbf{\Psi}$ and $\mathbf{S}$,
\begin{align}
&\frac1N\mathbf{S \Psi S}^\dagger = \frac1N \sum_{i = 1}^nN^{\alpha/2}R^{-\alpha} P_i \mathbf{g}_i\mathbf{g}_i^\dagger \nonumber \\
&\;\;\;\;\;\;\;\;\;\;\;\;\;\;\;\;\;\;\;\;\;\;\;\;\;\;\;\;+ \frac1N \sum_{i = 1}^nN^{\alpha/2}(r_i^{-\alpha}-R^{-\alpha}) P_i \mathbf{g}_i\mathbf{g}_i^\dagger\nonumber \\
& = \frac{N^{\frac{\alpha}{2}-1}}{R^{\alpha}}\sum_{i \in \mathcal{T}} \mathbf{g}_i\mathbf{g}_i^\dagger + N^{\frac{\alpha}{2}-1} \sum_{i \in\mathcal{T}} (r_i^{-\alpha}-R^{-\alpha})  \mathbf{g}_i\mathbf{g}_i^\dagger\,.\label{Eqn:MinEvalEquation1}
\end{align}
Note that $ \lim_{n\to\infty} \frac{|\mathcal{T}|}{n} = \nu$,in probability from the AIP. Recall that $\nu$ represents the fraction of potential interferers that are active in the limit. Let the members of $\mathcal{T}$ be labeled $i_1, i_2, \cdots $. Next, define a new set of indices $\mathcal{T}'$ which contains the indices
of at most $\frac{\nu n}{\delta}$ of the active nodes. In other words,
\begin{align}
\mathcal{T}' = \left\{i_k : k \leq \min\left(\frac{\nu n}{\delta}, |\mathcal{T}|\right)\right\}
\end{align}
Since $c = n/N$ and $n = \pi \rho_p R^2$,  $R = \sqrt{\frac{cN}{\pi \rho_p}}$. Thus, \eqref{Eqn:MinEvalEquation1} can be written as
\begin{align}
&\frac1N\mathbf{S \Psi S}^\dagger = \underbrace{\left(\frac{\pi \rho_p}{c}\right)^{\frac{\alpha}{2}}\frac{c|\mathcal{T}'|}{n} \frac{1}{|\mathcal{T}'|}\sum_{i \in \mathcal{T}'} \mathbf{g}_i\mathbf{g}_i^\dagger}_{\mathbf{T}_1}  \nonumber\\
 &+  \underbrace{\frac{N^{\frac{\alpha}{2}-1}}{R^{\alpha}} \sum_{i \in \mathcal{T} \backslash \mathcal{T}'} \mathbf{g}_i\mathbf{g}_i^\dagger + N^{\frac{\alpha}{2}-1} \sum_{i \in\mathcal{T}} (r_i^{-\alpha}-R^{-\alpha})  \mathbf{g}_i\mathbf{g}_i^\dagger }_{\mathbf{T}_2}\,.\notag 
\end{align}
Since $\mathbf{T}_2$ is non-negative definite, $\lambda_{min}(\mathbf{T}_2) \geq 0$. Thus, by the Weyl inequality,
\begin{align}
\lambda_{min}\left(\frac1N\mathbf{S \Psi S}^\dagger\right)  \geq \lambda_{min}(\mathbf{T}_1)+ \lambda_{min}(\mathbf{T}_2)\geq \lambda_{min}(\mathbf{T}_1)\,.\label{Eqn:Weyl}
\end{align}

Conditioned on $\mathcal{D}$, $\frac{|\mathcal{T}'|}{n} = \frac{\nu}{\delta}$, and from Theorem 5.11 of \cite{bai2009spectral},  with probability 1,
\begin{align}
\lambda_{min}\left(\frac{1}{|\mathcal{T}'|}\sum_{i \in \mathcal{T}'} \mathbf{g}_i\mathbf{g}_i^\dagger\right) \to \left(1-\sqrt{\frac{\delta}{c\nu}}\right)^2\,,
\end{align}
where recall that $1<\delta<c\nu$. From Theorem 6.3 of \cite{bai2009spectral}, $\exists n_0$ such that $\forall n > n_0$, with probability 1,
\begin{align}
\lambda_{min}\left(\frac{1}{|\mathcal{T}'|}\sum_{i \in \mathcal{T}'} \mathbf{g}_i\mathbf{g}_i^\dagger\right) \geq \left(1-\sqrt{\frac{\delta}{c\nu}}\right)^2\,,
\end{align}
From the definition of $\mathbf{T}_1$, and
\eqref{Eqn:Weyl},  we have  that conditioned on $\mathcal{D}$, there exists an $n_0$ for which for  all $n > n_0$, with probability 1,
\begin{align}
\lambda_{min}\left(\frac1N\mathbf{S \Psi S}^\dagger\right) \geq \lambda_{min}(\mathbf{T}_1) \geq  \left(\frac{\pi \rho_p}{c}\right)^{\frac{\alpha}{2}}\frac{c\nu}{\delta} \left(1-\sqrt{\frac{\delta}{c\nu}}\right)^2\notag
\end{align}
Dividing the RHS by 2 makes the inequality strict.

\subsection{Proof that the HC-I and HC-II models satisfy the asymptotic independence property} \label{Sec:HardCoreProof}

\begin{figure}[htpb]
\center
\includegraphics[width = 3in]{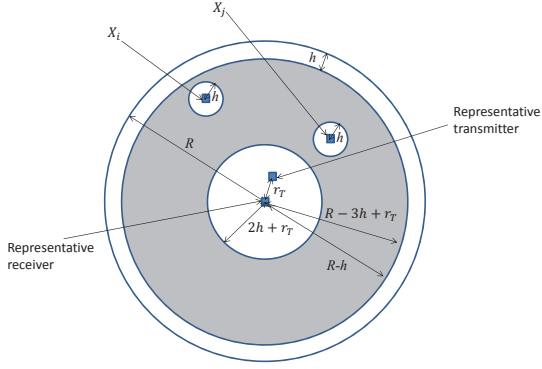}
\caption{Hard-core network illustrating the representative
receiver, representative transmitter and two interferers.}
\label{Fig:HC1Fig}
\end{figure}

In this appendix, we show that both the HC-I and HC-II models satisfy
\eqref{Eqn:AsympIndepSym}. Instead of showing that the
joint CDFs factor in the limit, it will be simpler to show that the
joint Complementary CDFs (i.e. 1-CDF) factor instead, which implies that the CDFs factor for two variables. Figure \ref{Fig:HC1Fig} illustrates the network with the
representative receiver at the origin, the representative transmitter
at a distance $r_T$ from the origin, and interferers
$i$ and $j$. Let $A$ denote the annulus centered at the origin
with inner radius $r_T+2h$, and outer radius $R - h$. Let $\mathcal{A}$ represent
the event that $X_i \in A, X_j \in A$ and $|X_i - X_j| > 2h$. We
consider this event in to handle edge effects, the effects of the representative transmitter and to limit the dependence between $P_i$ and $P_j$.
Note that conditioned on $\mathcal{A}$ and for finite $R$, $P_i$ and $P_j$ are dependent because there are $n$ points in the circular network.  For the rest of this derivation, we shall assume that $R > r_T + 5h$  which simplifies the subsequent analysis and does not change the final result since we take $R\to\infty$..
As $R \rightarrow\infty, \Pr(\mathcal{A})\to 1$ since  $\Pr(X_i\in A, X_j \in A)\to 1$  as the fraction of the area of the circle of
radius $R$  that is not in $A$ goes to zero, and  $\Pr\left(|X_i - X_j| \leq 2h\right)= \Theta(\frac{1}{R^2})\to 0$ (e.g. see \cite{Mathai} Equation 2.3.63). Since $x \geq 0$, $\Pr(P_i N^{\alpha/2} r_i^{-\alpha} > x, P_j N^{\alpha/2} r_j^{-\alpha} > x) > 0$ only if $P_i = P_j = 1$. Since $\Pr(\mathcal{A})\to 1$ as $R\to\infty$,
\begin{align}
&\Pr(P_i N^{\frac{\alpha}{2}} r_i^{-\alpha} > x, P_j N^{\frac{\alpha}{2}} r_j^{-\alpha} > x)\nonumber \\
&=\Pr(N^{\frac{\alpha}{2}} r_i^{-\alpha} > x, N^{\frac{\alpha}{2}} r_j^{-\alpha} > x|\mathcal{A}, P_i = 1, P_j = 1)\, \times \nonumber \\
&\Pr(P_i = 1, P_j = 1|\mathcal{A}).\label{eqn44}
\end{align}
The RHS of \eqref{eqn44} can be evaluated using the following two lemmas.
\begin{lemma} \label{Lemma:SIRConvergenceLemma1}
For both the HC-I and HC-II models, and $i \neq j$, we have
\begin{align}
&\lim_{n\to\infty}\Pr(N^{\frac{\alpha}{2}}r_i^{-\alpha}>x,N^{\frac{\alpha}{2}}r_j^{-\alpha}>x|\mathcal{A},
P_i=1,P_j=1) \nonumber \\ &\;\;\;\;\;\;\;\;=\lim_{n\to\infty}\Pr(N^{\frac{\alpha}{2}}r_j^{-\alpha}>x)\lim_{n\to\infty}\Pr(N^{\frac{\alpha}{2}}r_i^{-\alpha}>x)\, \;\;\;\;\;\mbox{and,}\label{eqn45}\\
&\lim_{n\to\infty}\Pr(N^{\frac{\alpha}{2}}r_i^{-\alpha}>x,|\mathcal{A}, P_i =1)=\lim_{n\to\infty}\Pr(N^{\frac{\alpha}{2}}r_j^{-\alpha}>x) \label{Eqn:LimitCond}
\end{align}
\emph{Proof: Given in Appendix E}
\end{lemma}
\begin{lemma} \label{Lemma:SIRConvergenceLemma2}
For both the HC-I and HC-II models, and $i \neq j$,
\begin{align}
\lim_{n\to\infty}&\Pr(P_i=1, P_j=1|\mathcal{A})=\nonumber \\
&\lim_{n\to\infty}\Pr(P_i=1)\lim_{n\to\infty}\Pr(P_j=1)\label{eqn46}
\end{align}
where $\lim_{n\to\infty}\Pr(P_i=1)=e^{-\pi\rho_ph^2}$
and $\lim_{n\to\infty}\Pr(P_i=1)=\frac{1-e^{-\pi\rho_ph^2}}{\pi\rho_ph^2
}$ for the HC-I and HC-II models respectively, for which $\nu$ from \eqref{Eqn:NuDef} follows directly.
\end{lemma}
\noindent\emph{Proof: Given in Appendix F}

Substituting \eqref{eqn46} and \eqref{eqn45} into \eqref{eqn44}, we have for $i \neq j$,
\begin{align}
&\lim_{n\to\infty}\Pr(P_i N^{\frac{\alpha}{2}}r_i^{-\alpha}>x, P_j
N^{\frac{\alpha}{2}}r_j^{-\alpha}>x)\nonumber \\
&=\lim_{n\to\infty}\Pr(P_i=1) \lim_{n\to\infty}\Pr(P_j=1) \lim_{n\to\infty}\Pr(N^{\frac{\alpha}{2}}r_j^{-\alpha}>x)\, \times \nonumber \\ &\;\;\;\;\;\;\;\;\;\;\lim_{n\to\infty}\Pr(N^{\frac{\alpha}{2}}r_i^{-\alpha}>x)\nonumber \\
&=\lim_{n\to\infty} \Pr(P_j
N^{\frac{\alpha}{2}}r_j^{-\alpha}>x) \lim_{n \rightarrow\infty}
\Pr(P_iN^{\frac{\alpha}{2}}r_i^{-\alpha}>x), \label{Eqn:HCLimitFactor}
\end{align}
where \eqref{Eqn:HCLimitFactor} is from  \eqref{Eqn:LimitCond}  and the fact that $\Pr(P_i = 1|\mathcal{A}) \to \Pr(P_i = 1)$, which also implies \eqref{Eqn:DistanceFactor}. Recalling the definition of  $p_{in}$, \eqref{Eqn:HCLimitFactor} implies \eqref{Eqn:AsympIndepSym}.


\subsection{Proof of Lemma
\ref{Lemma:SIRConvergenceLemma1}\label{Sec:ProofOfSmallestEval3}}
Recall that $\mathcal{A}$ is the event that $|X_i-X_j| > 2h$ and $X_i, X_j \in A$. Thus,
\begin{align}
&\Pr(N^{\frac{\alpha}{2}}r_i^{-\alpha}> x,
N^{\frac{\alpha}{2}}r_j^{-\alpha}> x|\mathcal{A}, P_i=1, P_j=1)\nonumber \\ 
&=\Pr(N^{\frac{\alpha}{2}}r_i^{-\alpha}>
x|\mathcal{A},N^{\frac{\alpha}{2}}r_j^{-\alpha}>
x)\Pr(N^{\frac{\alpha}{2}}r_j^{-\alpha}>
x|\mathcal{A}).\label{eqn48}
\end{align}
Let $|A| = \pi(R-h)^2-\pi(r_T+2h)^2$ be the area of $A$. Noting that $\mathcal{A}$ implies that  $X_i \notin B(X_j,2h)$,
\begin{align}
&\Pr(N^{\frac{\alpha}{2}}r_i^{-\alpha}>
x|\mathcal{A},N^{\frac{\alpha}{2}}r_j^{-\alpha}>
x)\nonumber \\
&=\Pr(r_i<\sqrt{N}x^{-\frac{1}{\alpha}}|\mathcal{A}, r_j
<\sqrt{N}x^{-\frac{1}{\alpha}})\notag\\
\hspace*{-1pc}&\leq\left(\frac{\pi N x^{{-\frac{2}{\alpha}}}}{|A|-4\pi h^2}
\right) 1_{\left\{N^{\frac{\alpha}{2}} (R-h)^{-\alpha}\leq x\leq
N^{\frac{\alpha}{2}}
(r_T+2h)^{-\alpha}\right\}}+\nonumber \\
&\;\;\;1_{\left\{x<N^{\frac{\alpha}{2}}
(R-h)^{-\alpha}\right\}}\label{Eqn:HCUB}
\end{align}
Similarly, we can write a lower bound which is active if $B(X_j, 2h) \subset B(0, \sqrt{N}
x^{-\frac{1}{\alpha}})$ as follows
\begin{align*}
\Pr(&N^{\frac{\alpha}{2}}r_i^{-\alpha}>
x|\mathcal{A}, N^{\frac{\alpha}{2}}r_j^{-\alpha}> x)=\nonumber \\
& \Pr(r_i <
\sqrt{N}x^{-\frac{1}{\alpha}}|\mathcal{A}, r_j < \sqrt{N}
x^{-\frac{1}{\alpha}})\\
&\geq\left(\frac{\pi N x^{{-\frac{2}{\alpha}}}-\pi(r_T+2h)^2 - 4\pi h^2}{|A|-4\pi h^2}
\right)\,\times \nonumber \\
& 1_{\left\{N^{\frac{\alpha}{2}} (R-h)^{-\alpha}\leq x\leq
N^{\frac{\alpha}{2}}
(r_T+2h)^{-\alpha}\right\}}+1_{\left\{x<N^{\frac{\alpha}{2}}
(R-h)^{-\alpha}\right\}}
\end{align*}
Comparing with \eqref{Eqn:HCUB} and recalling that $|A| = \Theta(R^2) = \Theta(N)$, note that the lower bound converges to the same form as the upper bound as $n,N, R\to\infty$. Thus, taking the limit of  \eqref{Eqn:HCUB} and using  $n/N = c$ and \eqref{Eqn:MaternPotentialInterferers} yields
\begin{align}
&\lim_{n
\rightarrow\infty}\Pr(N^{\frac{\alpha}{2}}r_i^{-\alpha}>
x|\mathcal{A}, N^{\frac{\alpha}{2}}r_j^{-\alpha}> x)\nonumber \\
&=
\left(\frac{\pi\rho}{c}\right)x^{-\frac{2}{\alpha}} 1_{\left\{\left(\frac{\pi\rho_p}{c}\right)^{a/2}\leq
x\right\}}+1_{{\left\{x<\left(\frac{\pi\rho_p}{c}\right)^{a/2}\right\}}}.\label{eqn49}
\end{align}
Since without conditioning on $\mathcal{A}$, $X_i$ and $X_j$ are i.i.d., and $\Pr(\mathcal{A}) \to 1$, from \eqref{eqn49},
\begin{align}
&\lim_{n \rightarrow\infty}\Pr(N^{\frac{\alpha}{2}}r_j^{-\alpha}>
x|\mathcal{A})= \lim_{n
\rightarrow\infty}\Pr(N^{\frac{\alpha}{2}}r_i^{-\alpha}>
x)=\nonumber \\
&\left(\frac{\pi\rho}{c}\right)x^{-\frac{2}{\alpha}} 1_{\left\{\left(\frac{\pi\rho_p}{c}\right)^{a/2}\leq
x\right\}}+1_{{\left\{x<\left(\frac{\pi\rho_p}{c}\right)^{a/2}\right\}}}. \label{Eqn:Tmp}
\end{align}
Taking the limit of \eqref{eqn48}, and
substituting \eqref{eqn49} and \eqref{Eqn:Tmp} yields \eqref{eqn45}. \eqref{Eqn:LimitCond} follows directly since conditioned  on $\mathcal{A}$, $P_i$ is independent of $X_i$, and hence $r_i$, and moreover $\Pr(\mathcal{A})\to 1$.

\subsection{Proof of Lemma \ref{Lemma:SIRConvergenceLemma2}}
Let \#$B(X_i, h) = |\{\ell:\ell\neq i, X_\ell \in B (X_i, h)\}|$
denote the number of potential interferers in $B(X_i,h)$ excluding the interferer at $X_i$ itself. For both
the HC-I and HC-II models, conditioned on \#$B(X_i, h) =
k$ and $\mbox{\#}B(X_j, h) = m$, the events $P_i = 1$ and $P_j = 1$ are
independent. Thus,
\begin{align}
&\Pr(P_i=1, P_j=1,\mbox{\#}B(X_i, h)=k, \mbox{\#}B(X_j,
h)=m|\mathcal{A})=\nonumber \\
&\Pr(\mbox{\#}B(X_i, h)=k, \mbox{\#}B(X_j,
h)=m|\mathcal{A}) \nonumber\\
&\;\;\;\;\;\;\;\;\;\;\;\times\Pr(P_i=1|\mathcal{A},\mbox{\#}B(X_i,
h)=k)\nonumber\\
&\;\;\;\;\;\;\;\;\;\;\;\times\Pr(P_j=1|\mathcal{A},\mbox{\#}B(X_j, h)=m).\label{eqn56}
\end{align}
Summing \eqref{eqn56} over all $m$ and $k$ excluding the potential interferers at $X_i$ and $X_j$ yields,
\begin{align}
&\Pr(P_i=1,
P_j=1|\mathcal{A})=\nonumber \\
&\sum^{n-2}_{k=0}\sum^{n-2}_{m=0}\Pr(P_i=1|\mathcal{A},\mbox{\#}B(X_i,
h)=k)\,\times \nonumber \\
&\Pr(P_j=1|\mathcal{A}, \mbox{\#}B(X_j, h)=m)\,\times \nonumber \\
&\Pr(\mbox{\#}B(X_i, h)=k, \mbox{\#}B(X_j,
h)=m|\mathcal{A}).\label{eqn57}
\end{align}

{\bf For the HC-I model}, $P_i = 1$ and $P_j = 1$ if and only if $\mbox{\#}B(X_i, h) = 0$ and $\mbox{\#}B(X_j, h) = 0$, respectively. Hence \eqref{eqn57} becomes
\begin{align}
&\Pr(P_i=1,
P_j=1|\mathcal{A}) =  \Pr(\mbox{\#}B(X_j, h)=0, \mathcal{A}) \times\nonumber \\
&\;\;\;\;\;\;\;\;\Pr(\mbox{\#}B(X_i, h)=0|\mathcal{A}, \mbox{\#}B(X_j, h)=0)\, \label{eqn58}
\end{align}
Recall that conditioned on $\mathcal{A}$, $X_i\notin B(X_j,h)$, and $P_j = 1$ if the $n-2$ remaining potential interferers are not in $B(X_j,h)$.  Using \eqref{Eqn:MaternPotentialInterferers} and $\Pr(\mathcal{A})\to1$ as $n,N,R\to\infty$,
\begin{align}
&\lim_{n\to\infty}\Pr(\mbox{\#}B(X_j, h)=0|\mathcal{A})= \lim_{n\to\infty}\Pr(P_j=1) = \nonumber\\
 &\lim_{R\to\infty} \!\left(1-\frac{\pi h^2}{\pi R^2-\pi(r_t+2h)^2}\right)^{\pi\rho_pR^2-2}\!=e^{-\pi\rho_ph^2}.\label{eqn60}
\end{align}
Conditioned on $\mbox{\#}B(X_j, h) = 0$ the $n-2$ other potential interferers are outside $B(X_j,h)$. Since $P_i = 1$ only if the remaining $n-2$ potential interferers are outside $B(X_i,h)$,
\begin{align}
&\Pr(\mbox{\#}B(X_i, h) = 0|\mathcal{A}, \mbox{\#}B(X_j, h) = 0) =\nonumber \\
&\left(1-\frac{\pi h^2}{\pi R^2-\pi(r_t+2h)^2-\pi h^2}\right)^{\pi\rho_pR^2-2}\,,\notag\\
&\lim_{n\to\infty}\Pr(\mbox{\#}B(X_i, h) = 0|\mathcal{A},
\mbox{\#}B(X_j, h) = 0)=e^{-\pi\rho_ph^2}\nonumber \\
&\;\;\;\;\;\;\;\;\;\;\;\;\;\;\;\; =
\lim_{n\to\infty}\Pr(P_i=1),\label{eqn61}
\end{align}
where the last equality follows from symmetry between $X_i$ and
$X_j$ and \eqref{eqn60}. Taking the limit of \eqref{eqn58} and
substituting \eqref{eqn60},
\eqref{eqn61} and $\Pr(\mathcal{A})\to 1$,  yields \eqref{eqn46} for the HC-I model.

{\bf For the HC-II model}, using the fact that $\Pr(P_i=1|\mbox{\#}B(X_i, h) = k, \mathcal{A}) = \frac{1}{k+1}$, \eqref{eqn57} becomes
\begin{align} &\Pr(P_i=1,
P_j=1|\mathcal{A})=\sum^{n-2}_{k=0}\sum^{n-2}_{m=0}\frac{1}{(k+1)(m+1)}\, \times\nonumber \\
&\Pr(\mbox{\#}B(X_i,
h) = k, \mbox{\#}B(X_j, h) = m|\mathcal{A})\notag\\
&=\sum^{n-2}_{k=0}\frac{1}{k+1}\Pr(\mbox{\#}B(X_i, h) =
k|\mathcal{A})\,\times \nonumber \\
&\sum^{n-k-2}_{m=0}\frac{1}{m+1}\Pr(\mbox{\#}B(X_j, h)
=
m|\mbox{\#}B(X_i, h) = k, \mathcal{A}),\notag
\\
&=\sum^{n-2}_{k=0}\frac{1}{k+1}{n-2\choose k}\left(\frac{\pi{h}^2}{|A|}\right)^k\left(1-\frac{\pi{h}^2}{|A|}\right)^{n-k-2} \times\notag\\
&\sum^{n-k-2}_{m=0}\frac{1}{m+1}{n-k-2\choose m}\left(\frac{\pi{h}^2}{|A|- \pi h^2}\right)^m \times \nonumber \\
&\;\;\;\;\;\;\left(1 - \frac{\pi{h}^2}{|A| -
\pi h^2}\right)^{n-k-2-m}\label{eqn64}
\end{align}
Consider the previous expression written in terms of a double series
(please see e.g. \cite{Habil} for basic properties of double series) with
the variables $n_1$ and $n_2$ replacing $n$, in the terms in the
summation, and the upper limit of the summation respectively.
Additionally, from \eqref{Eqn:MaternPotentialInterferers} and $|A| = \pi R^2 - \pi(r_T+2h)^2$,
\begin{align}
&\Pr(P_i=1, P_j=1|\mathcal{A})\nonumber \\
&=\sum^{n_2-2}_{k=0}\frac{1}{k+1}{n_1-2\choose k}\left(\frac{{h}^2}{\frac{n_1}{\rho_p}-\!(r_T+2h)^2}\right)^k\notag \\
&\;\;\;\;\;\times \left(1-\frac{{h}^2}{\frac{n_1}{\rho_p}-\!(r_T+2h)^2}\right)^{n_1-k-2}\notag\\
&\times\sum^{n_2-k-2}_{m=0}\frac{1}{m+1}{n_1-k-2\choose m }\left(\frac{{h}^2}{\frac{n_1}{\rho_p}-\!(r_T+2h)^2 - h^2}\right)^m\,
 \nonumber \\
&\;\;\times\left(1-\frac{{h}^2}{n_1/\rho_p - h^2}\right)^{n_1-k-2-m}\label{eqn65}
\end{align}
\noindent Taking the limits of \eqref{eqn65} as $n_1\to\infty$ followed by $n_2\to\infty$,
\begin{align}
&\lim_{n_2\to\infty}\lim_{n_1\to\infty}\!\Pr(P_i=1, P_j=1|\mathcal{A}) \nonumber \\
&=\lim_{n_2\to\infty}\sum^{n_2-2}_{k=0}\frac{(\pi\rho_bh^2)^ke^{-\pi{h}^2\rho_p}}{(k+1)!}
\sum^{n_2-k-2}_{m=0}\frac{(\pi\rho_bh^2)^me^{-\pi{h}^2\rho_p}}{(m+1)!}\label{eqn66a}
\end{align}
Since both the sums on the RHS of \eqref{eqn66a} are
finite for all $n_2$, by Theorem 2.11 of \cite{Habil},
\begin{align}
&\lim_{n\to\infty}\Pr(P_i=1, P_j=1|\mathcal{A}) =
\lim_{n_1,n_2\to\infty}\Pr(P_i=1, P_j=1|\mathcal{A})\notag\\
&=\lim_{n_2\to\infty}\lim_{n_1\to\infty}\Pr(P_i=1,
P_j=1|\mathcal{A}) =
\left(\frac{1-e^{-\pi{h}^2\rho_p}}{\pi\rho_bh^2}\right)^2.\label{eqn67}
\end{align}
Following similar steps, we have
\begin{align}
\lim_{n\to\infty}\Pr(P_i=1) = \lim_{n\to\infty}\Pr(P_j=1) =
\frac{1-e^{-\pi{h}^2\rho_p}}{\pi\rho_bh^2}.\label{eqn68}
\end{align}
Substituting \eqref{eqn68} into \eqref{eqn67} yields \eqref{eqn46}, completing the proof.

\subsection{Outline of Proof that Cellular Uplink Model Satisfies AIP} \label{Sec:FreqReuseMod}
Since this proof is similar to that of the HC-II model given above, only an summary of the main steps are given here.
Let $\mathcal{C}_A$ denote the union of all cells assigned to band-0, which are  wholly contained in $B(0,R)$, i.e., $\mathcal{C}_A = \bigcup_{k:\, \mathcal{C}_k \subset B(0,R)}\mathcal{C}_k$. Now consider the following
\begin{align}
& \Pr\left(P_i^{-\frac{1}{\alpha}} r_i \leq \sqrt{N}x^{-\frac{1}{\alpha}}| X_i \in\mathcal{C}_A,P_i = 1\right)\nonumber \\
&\;\;\;\;\;\;\;\;\;\;\;\;\;\;\;\;\;\; = \frac{1}{\left|\mathcal{C}_A\right|}\left|\mathcal{C}_A\cap B(0,\sqrt{N}x^{-\frac{1}{\alpha}})\right| \label{Eqn:NTDMADistBound}
\end{align}
 As $n,R\to\infty$, the edge effects diminish, and
\begin{align}
&\lim_{n\to\infty}\Pr\left(P_i^{-\frac{1}{\alpha}} r_i \leq \sqrt{N}x^{-\frac{1}{\alpha}}|X_i\in \mathcal{C}_A,P_i = 1\right)= \nonumber \\ &\lim_{n\to\infty}\frac{1}{\left|\mathcal{C}_A\right|}{\left|\mathcal{C}_A\cap B(0, \sqrt{N}x^{-\frac{1}{\alpha}})\right|} \nonumber \\
&=\left(\frac{\pi \rho}{c}x^{-\frac{2}{\alpha}}\right)1_{\left\{\left(\frac{\pi\rho_p}{c}\right)^{a/2}\leq
x\right\}} = \Pr(r_i \leq \sqrt{N}x^{-\frac{1}{\alpha}})\label{Eqn:NTDMADistLimit3}
\end{align}
Following steps used to derive \eqref{eqn67}, it can be shown that
\begin{align}
 &\lim_{n\to\infty}\Pr(P_i = 1, |X_i\in\mathcal{C}_A)=\frac{\rho_c}{\rho_p}{\left(1-e^{-\rho_p/\rho_c}\right)} \label{Eqn:NTDMASingleSumLimit}
\end{align}
Taking the product of \eqref{Eqn:NTDMADistLimit3} and \eqref{Eqn:NTDMASingleSumLimit}, and using the fact that $\Pr(X_i \in \mathcal{C}_A) \to \frac{1}{\kappa}$,
\begin{align}
&\lim_{n\to\infty}\Pr\left(P_ir_i^{-\alpha}N^{\frac{\alpha}{2}} \leq x\right)\nonumber \\
 &= \frac{1}{\kappa}\left(\frac{1-e^{-\rho_p/\rho_c}}{\rho_p/\rho_c}\right)\left(\frac{\pi \rho}{c}x^{-\frac{2}{\alpha}}\right)1_{\left\{\left(\frac{\pi\rho_p}{c}\right)^{a/2}\leq
x\right\}}
\end{align}
As $R\to\infty$, the probability that $X_i$ and $X_j$ are in the same cell $\to 0$, which implies \eqref{Eqn:AsympIndepSym}, since
\begin{align*}
&\lim_{n\to\infty}\Pr\left(P_i r_i^{-\alpha}N^{\frac{\alpha}{2}} >x,P_j r_j^{-\alpha}N^{\frac{\alpha}{2}} > x\right) \nonumber \\
 &= \lim_{n\to\infty}\Pr\left(P_i r_i^{-\alpha}N^{\frac{\alpha}{2}} >x\right)\,\lim_{n\to\infty}\Pr\left(P_j r_j^{-\alpha}N^{\frac{\alpha}{2}} > x\right)\,.
\end{align*}

\bibliography{IEEEabrv,main}

\end{document}